\documentclass[prd,aps,showpacs,preprintnumbers,amssymb]{revtex4}

\usepackage{epsf}

\def\e3p{$N N'$}

\begin{document}

\title{%
\hfill{\normalsize\vbox{%
\hbox{}
 }}\\
{
Mass Uncertainties of $f_0(600)$ and $f_0(1370)$ and
their Effects on Determination of the Quark and Glueball
Admixtures of the $I=0$ Scalar Mesons
}}                    

\author{Amir H. Fariborz \footnote[1]{Email:
fariboa@sunyit.edu}}

\affiliation{Department of Mathematics/Science,
 State University of New York Institute of Technology, Utica,
 NY 13504-3050, USA\\\\
}
                                                      
\date{\today}

\begin{abstract}

Within a nonlinear chiral Lagrangian framework the
correlations between the quark and glueball admixtures of
the isosinglet scalar mesons below 2 GeV and the current
large uncertainties on the mass of the $f_0(600)$ and the
$f_0(1370)$ are studied.  The framework is formulated in
terms of two scalar meson nonets (a two-quark nonet and a
four-quark nonet)  together with a scalar glueball.  It
is shown that while some properties of these states are
sensitive to the mass of $f_0(600)$ and $f_0(1370)$,
several relatively robust conclusions can be made:  The
$f_0(600)$, the $f_0(980)$, and the $f_0(1370)$ are
admixtures of two and four quark components, with
$f_0(600)$ being dominantly a non-strange four-quark
state, and $f_0(980)$ and $f_0(1370)$ having a dominant
two-quark component.  Similarly, the $f_0(1500)$ and the
$f_0(1710)$ have considerable two and four quark
admixtures, but in addition have a large glueball
component.  For each state, a detailed analysis providing
the numerical estimates of all components is given. It is
also shown that this framework clearly favors the
experimental values: $m^{\rm exp.} [f_0(600)] < 700$ MeV
and $m^{\rm exp.} [f_0(1370)]$ = 1300$-$1450 MeV.  
Moreover, an overall fit to the available data shows a
reciprocal substructure for the $f_0(600)$ and the
$f_0(1370)$, and a linear correlation between their
masses of the form $m [f_0(1370)] = 0.29 \, m [f_0(600)]
+ 1.22 $ GeV. The scalar glueball mass of 1.5$-$1.7 GeV
is found in this analysis.

\end{abstract}


\maketitle


\section{Introduction}


Exploring the properties of the scalar mesons is known to
be a non trivial task in low energy QCD.  This is due to
issues such as their large decay widths and overlap with
the background, as well as having several decay channels
over a tight energy range \cite{PDG}.  Particularly, the case of
$I=0$ states is even more complex due to their various
mixings, for example, among two and four quark states and
scalar glueballs.
Below 1 GeV, the well-established experimental states
are \cite{PDG}: the $f_0(980)$ [$I=0$] and the $a_0(980)$
[$I=1$], together with states that have uncertain
properties: the $f_0(600)$ or $\sigma$ [$I=0$] with a mass of 
400$-$1200 MeV, and a decay width of 600$-$1000 MeV, and the 
$K_0^*(800)$ or $\kappa$ [$I=1/2$]
which is not yet listed but cautiously discussed in PDG.  
The
existence of the $K_0^*(800)$ has been confirmed in some
theoretical models, while has been disputed in
some other approaches.
In the range of 1$-$2 GeV, the listed scalar
states are \cite{PDG}: $K_0^*(1430)$ [$I=1/2$];
$a_0(1450)$ [$I=1$]; and $f_0(1370)$, $f_0(1500)$,
$f_0(1710)$ [$I=0$].  The isodoublet and isotriplet
states, are generally believed to be closer to $q {\bar
q}$ objects, even though some of their properties
significantly deviate from such a description. Among the
heavier isosinglet states, the $f_0(1370)$ has the
largest experimental uncertainty \cite{PDG} on its mass
(1200$-$1500 MeV) and decay width (200$-$500 MeV), 
and other states in the energy
range of around 1.5 GeV (or above) seem to contain a
large glue component and maybe  good candidates for the
lowest scalar glueball state.

A simple quark-antiquark description is known to fail
for the lowest-lying scalar states,  and that has made 
them the focus of intense investigation for a long
time.
Several foundational scenarios for their substructures 
have been considered, including MIT bag model \cite{Jaf}, $K {\bar
K}$ molecule \cite{Isg} and unitarized quark model
\cite{Tor}, and many theoretical frameworks for  
the properties of the scalars have been developed, including, among 
others, chiral Lagrangian of refs. 
\cite{San,BFSS1,BFSS2,Far,pieta,Mec,LsM,02_BHS,Higgs,e3p,04_F1,04_F2,05_FJS}, 
upon which the 
present investigation is based.   Many recent works 
\cite{Eli,00_CK,02_CT,02_TKM,04_TKM,04_NR,05_GGF,05_GGLF,05_VVFS,05_BNNB,05_N,06_MPPR}
have investigated 
different aspects of the scalar mesons, particularly, their 
family connections and possible description in terms of meson 
nonet(s), which proved a comparison with this work.

Various ways of grouping the scalars together have been
considered in the literature.  For example, in some
approaches, the properties of the scalars above 1 GeV
(independent of the states below 1 GeV) are investigated,
whereas  other works have only focused on the 
states
below 1 GeV.  There are also works that have investigated
several possibilities for grouping together some of the
states above 1 GeV with those below 1 GeV.

In addition to various ways that the physical states may be grouped 
together,
different bases out of quark-antiquark, four quark, and glueball have been
considered for their internal structure.  For 
example,
in a number of investigations the $I=0$ states above 1 GeV are studied
within a framework which incorporates quark-antiquark and glueball
components.
However, in lack
of a complete framework for understanding the properties
of the scalar mesons it seems more objective to develop
general frameworks in which, a priori, no specific
substructure for the scalars is assumed, and instead, all
possible components (quark-antiquark, four-quark, and
glueball) are considered.  The framework of the
present work considers all such components for the $I=0$
states, and studies the five listed $I=0$ states below 2 GeV \cite{PDG}.

Besides the generality of the framework, there are supportive indications 
that the 
lowest 
and 
the next-to-lowest lying
scalar states have admixtures of quark-antiquark and four-quark 
components.
For example, it is
shown in \cite{Mec} that the $a_0(1450)$ and the
$K_0^*(1430)$ have a considerable two and four quark
admixtures, which provides a basis for explaining the
mass spectrum and the partial decay widths of the $I=1/2$
and $I=1$ scalars [$K_0^*(800), a_0(980), K_0^*(1430)$
and $a_0(1450)$].    It then raises the question that if the $I=1/2$ and
$I=1$ scalars below 2 GeV are admixtures of two and four quark components,
why should not the $I=0$ states be blurred with such a mixing 
complexity?     Answering this 
question is the main motivation of this paper, and we will see that this 
framework shows that there is a substantial admixture of the two and 
the four 
quark components for the $I=0$ scalars below 2 GeV.

Motivated by the importance of the
mixing for the $I=1/2$ and $I=1$ in this framework,
the case of the $I=0$ states was initially studied 
in
\cite{04_F1,04_F2} in which mixing with a scalar glueball is also
included.  In \cite{04_F1} the parameter space of the
$I=0$ Lagrangian, which is already constrained by the properties of the 
$I=1/2$ and $I=1$ states in ref. \cite{Mec},  is studied using the mass 
spectrum 
and several two-pseudoscalar decay widths and decay ratios of the $I=0$
states.  The present work extends the work of \cite{04_F1} by
investigating in detail the effect of the 
mass uncertainties
of the $f_0(600)$ and $f_0(1370)$ on the components
of the $I=0$ states below 2 GeV.   It also
provides an insight into the likelihood of the mass of
$f_0(600)$ and $f_0(1370)$ within their experimental values \cite{PDG} 
in ranges 400-1200 MeV and
1200-1500 MeV, respectively.  In addition a linear
correlation between these two masses is predicted by the
model.

\begin{figure}[h]
\epsfxsize=10cm
\epsfbox{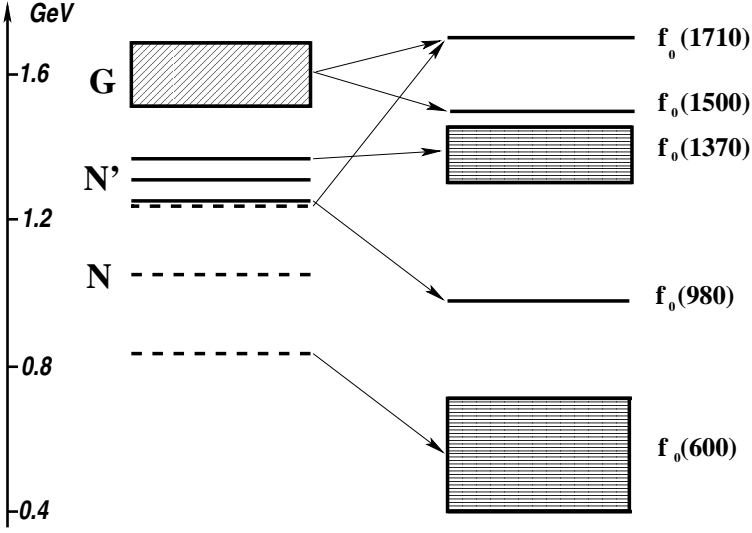}
\vspace*{8pt}
\caption{
	Prediction of the present model for the formation of the $I=0$
scalar states below 2 GeV.  On the right, the physical states are shown,
with the solid lines representing the experimentally well established
masses, and the height of the two boxes representing the prediction of the
present model for the uncertainties of the mass of $f_0(600)$ and
$f_0(1370)$, which are approximately in ranges 0.4 - 0.7 GeV and 1.3 -
1.45 GeV, respectively.  On the left, the ``bare'' states are shown, with
the dashed lines representing the four-quark nonet $N$, the solid lines
representing the two-quark nonet $N'$ and the box representing the scalar
glueball.  The mass of the bare states in the two nonets (from bottom to
top) are:  $m({\bar u}{\bar d} u d)$ = 0.83 GeV, $m({\bar d}{\bar s} u d)$
= 1.06 GeV, $m({\bar s}{\bar d} d s + {\bar s}{\bar u} u s)  
/\sqrt{2})=m({\bar u} u + {\bar d} d)/\sqrt{2})$ = 1.24 GeV, $m({\bar u}
s)$ = 1.31 GeV and $m({\bar s}s)$ = 1.38 GeV.  The uncertainty of the
glueball mass (shown by the height of the box, approximately between 1.5
GeV to 1.7 GeV) is due to the uncertainty of the mass of $f_0(600)$ and
$f_0(1370)$.  The arrows show the dominant component(s)  of each physical
state.  More details are given in Figs.  \ref{F_comps_vs_ms} and
\ref{F_comps_error}.
 \label{F_mixing}
}
\end{figure}

Specifically, we will numerically analyze the mass spectrum and perform an 
inverse problem:   knowing the mass of the physical states (or in the 
case of the $f_0(600)$ and the $f_0(1370)$ a wide experimental range  
for their masses) we search for the parameter space of the 
Lagrangian (which is formulated in terms of a two quark nonet and a 
four quark nonet) and find solutions that can reproduce these masses.    
Once the parameter space is determined it provides information about the 
properties of the two nonets.   The results are summarized in 
Fig.\ref{F_mixing}.
We will show how the present model describes the $I=0$ scalar states in
terms of a four-quark nonet $N$ which lies in the range of 0.83-1.24 GeV,
together with a two-quark nonet $N'$ in the 1.24-1.38 GeV range, and a
scalar glueball that this model predicts in the range of 1.5-1.7 GeV
(figure \ref{F_mixing}).  The mass range of the two quark scalar nonet 
$N'$ is qualitatively consistent with the expected range of 1.2 GeV from 
spectroscopy of $p$-wave mesons.    We will see that this analysis shows 
that, similar to the case of $I=1/2$ and $I=1$ scalar mesons,
the $I=0$ scalars have a significant admixtures of two and four quark 
components, and in addition the $f_0(1500)$ and the $f_0(1710)$ have a 
dominant glueball content.       The dominant component(s) of each state
is summarized in Fig. \ref{F_mixing}.

The underlying framework of the present work has been previously applied 
in analyzing  numerous low-energy processes that involve scalar mesons,  
and 
consistent pictures have emerged.    The 
existence of the $f_0(600)$ (or $\sigma$) and the $K_0^*(800)$ (or 
$\kappa$) and 
their properties have been investigated in refs. \cite{San,BFSS1}. The 
lowest-lying nonet of scalars and a four-quark interpretation of these 
states is studied in \cite{BFSS2} and applied to $\pi\eta$ 
scattering in \cite{pieta} and several decays such as 
$\eta'\rightarrow\eta\pi\pi$ \cite{Far} and $\eta\rightarrow 3\pi$ 
\cite{e3p} and radiative $\phi$ decays \cite{02_BHS}.     In addition to 
ref. 
\cite{Mec}, the mixing between a 
two quark and a four quark nonets is also investigated in \cite{LsM} and 
\cite{05_FJS}.

We describe the theoretical framework in Sec. 1,
followed by the numerical results in Sec.2, and a short summary in Sec. 3.

\section{Theoretical Framework}


\subsection{Mixing Mechanism for I=1/2 and I=1 Scalar 
States}

In ref. \cite{Mec} the properties of the $I=1/2$ and
$I=1$ scalar mesons, $\kappa(900)$, $K_0^*(1430)$,
$a_0(980)$ and $a_0(1450)$, in a nonlinear chiral
Lagrangian framework is studied in detail. In this
approach, a ${\bar q} {\bar q} q q$ nonet $N$ mixes with
a ${\bar q} q$ nonet $N'$ and provides a description of
the mass spectrum and decay widths of these scalars.  
The $K_0^*(1430)$ and the $a_0(1450)$,
are generally believed to be good candidates for a
${\bar q} q$ nonet \cite{PDG}, but some of their
properties do not quite follow this scenario. For
example, in a $q {\bar q}$ nonet, isotriplet is expected
to be lighter than the isodoublet, but for these two
states \cite{PDG}:
\begin{equation}
m \left[ a_0(1450) \right] = 1474 \pm 19 \hskip .2cm {\rm MeV} > m 
\left[ K_0^*(1430) \right] =
1412 \pm 6  \hskip .2cm {\rm MeV}
\label{a0k0_mass_exp}
\end{equation}
Also their decay ratios given in PDG \cite{PDG} do not follow a pattern 
expected from an SU(3) symmetry (given in parenthesis):
\begin{eqnarray}
{ {\Gamma \left[ a_0^{total} \right]}\over
{\Gamma\left[K_0^*\rightarrow\pi K\right] }} &=& 0.92 \pm 0.12 
\hskip .2cm (1.51)\nonumber \\
\hskip 0.2cm
{ {\Gamma \left[ a_0\rightarrow K{\bar K} \right]}\over
{\Gamma\left[a_0\rightarrow\pi \eta\right] }} &=& 0.88 \pm 0.23 
\hskip .2cm  (0.55)\nonumber \\
\hskip 0.2cm
{ {\Gamma \left[ a_0\rightarrow \pi\eta' \right]}\over
{\Gamma\left[a_0\rightarrow\pi \eta\right] }} &=& 0.35\pm0.16 \hskip 
.2cm  (0.16)
\label{a0k0_decay_exp}
\end{eqnarray}
These properties of the $K_0^*(1430)$ and the $a_0(1450)$ are naturally  
explained by the mixing mechanism of ref. \cite{Mec}.   
The general mass terms for  the $I=1/2$ and the $I=1$ states 
can be written as
\begin{equation}
{\cal L}_{mass}^{I=1/2,1} = 
- a {\rm Tr}(NN) - b {\rm Tr}(NN{\cal M}) 
- a' {\rm Tr}(N'N') - b' {\rm Tr}(N'N'{\cal M}) 
\label{L_mass_I1}
\end{equation}
where ${\cal M} = {\rm diag} (1,1,x)$ with $x$ being the ratio of
the strange to non-strange quark masses, and $a,b,a'$ and $b'$ are 
unknown parameters fixed by the unmixed or ``bare'' 
masses (denoted below by subscript ``0''):
\begin{equation}
\begin{array}{cclcccl}
m^2 [a_0] &=& 2 (a + b) & &  m^2 [a'_0] & = & 2 (a' + b')\\
m^2 [K_0] &=& 2a + (1+x) b & &  m^2 [K'_0] & = & 2 a' + (1+x) b'
\end{array}
\end{equation}
As $N$ is a 
four-quark nonet and $N'$ a two-quark nonet, we expect:
\begin{equation}
m^2 [K_0] <  m^2 [a_0] \le m^2 [a'_0] < m^2 [K'_0] 
\label{mass_order}
\end{equation}
In fact, this is how we tag $N$ and $N'$ to a four-quark and 
a two-quark nonet, respectively.  Introducing a general mixing 
\begin{equation}
{\cal L}_{mix}^{I=1/2,1} = 
-\gamma {\rm Tr} \left( N N' \right) 
\label{L_mix_I1}
\end{equation}
it is shown in \cite{Mec} that  for
$0.51 < \gamma < 0.62 \hskip .2cm{\rm GeV^2}$, it is possible to 
recover the physical masses such that the ``bare'' masses have the 
expected ordering of (\ref{mass_order}).
In this mechanism, the ``bare'' isotriplet states split more 
than the isodoublets, and 
consequently, the
physical isovector state $a_0(1450)$ becomes heavier than the  
isodoublet state $K_0^*(1430)$  in agreement with the observed 
experimental values in 
(\ref{a0k0_mass_exp}).  The 
light isovector and 
isodoublet states are the $a_0(980)$ and the
$K_0^*(800)$.    With the physical masses $m[a_0(980)]=0.9835$ GeV, 
$m [K_0^*(800)] =0.875$ GeV, $m[a_0(1450)]=1.455$ GeV and 
$m[K_0(1430)]=1.435$ GeV, the best values of mixing parameter $\gamma$ and 
the ``bare''
masses are found in \cite{Mec}
\begin{equation}
m_{a_0}=m_{a'_0}=1.24 \hskip .1cm{\rm GeV},\hskip .2cm 
m_{K_0}=1.06 \hskip .1cm{\rm GeV}, \hskip .2cm 
m_{K'_0}=1.31 \hskip .1cm{\rm GeV}, \hskip .2cm 
\gamma = 0.58 \hskip .1cm {\rm GeV^2}
\label{a0k0_m_bare}
\label{bare_masses}
\end{equation} 
These parameters are then used to study the decay widths of the $I=1/2$ 
and $I=1$ states \cite{Mec}.    A general Lagrangian describing 
the coupling of the two nonets $N$ and $N'$ to two-pseudoscalar particles 
are introduced and the unknown Lagrangian parameters are found by fits to 
various decay widths.

In summary, the work of ref. \cite{Mec} clearly shows
that properties of the lowest and the next-to-lowest
$I=1/2$ and $I=1$ scalar states can be described by a
mixing between a quark-antiquark nonet and a four quark
nonet.  It is concluded that the $I=1$ states are close
to maximal mixing (i.e. $a_0(980)$ and $a_0(1450)$ are
approximately made of 50\% quark-antiquark and 50\%
four-quark components), and the $I=1/2$ states have a
similar structure with $K_0^*(800)$ made of
approximately 74\% of four-quark and 26\%
quark-antiquark, and the reverse structure for the
$K_0^*(1430)$.  Now if the lowest and the next-to-lowest
$I=1/2$ and $I=1$ states have a substantial mixing of
quark-antiquark and four-quark components, then it seems
necessary to investigate a similar scenario for the $I=0$
scalars, which, in addition, can have a glueball
component as well.  In other words, if the lowest and the
next-to-lowest scalars are going to be grouped together, 
then all
components of quark-antiquark, four-quark and glueball 
for the $I=0$ states should be taken into account.  The case of $I=0$ 
states will be
discussed in next section.


\subsection{Isosinglet States}


The  $I=0$ scalars have been investigated in many recent works,  
including the general approach with two and four quark components 
as well as a glueball component of refs. \cite{04_F1,04_F2}.   Within the 
framework 
of 
ref. \cite{Mec}, an initial analysis is given in \cite{04_F1} in which the 
mass and the decay Lagrangian for the $I=0$ scalar mesons are studied in 
some details.  
The general mass terms for nonets $N$ and $N'$, and a scalar glueball $G$ 
can be 
written as:
\begin{eqnarray}
{\cal L}_{mass}^{I=0}  &=& {\cal L}_{mass}^{I=1/2,1} 
- c {\rm Tr}(N){\rm Tr}(N) 
- d {\rm Tr}(N) {\rm Tr}(N{\cal M}) 
\nonumber \\
&&- c' {\rm Tr}(N'){\rm Tr}(N') 
- d' {\rm Tr}(N') {\rm Tr}(N'{\cal M}) 
- g G^2
\label{L_mass_I0}
\end{eqnarray}
The unknown parameters $c$ and $d$ induce ``internal'' mixing 
between the two $I=0$ flavor combinations [$(N_1^1 + N_2^2)/\sqrt{2}$ and 
$N_3^3$] of nonet $N$. Similarly,  $c'$ and $d'$ play the same 
role in nonet $N'$.  
Parameters $c,d,c'$ and $d'$ do not contribute to 
the mass spectrum of the $I=1/2$ and $I=1$ states.
The last term represents the glueball mass term.  
The term ${\cal 
L}_{mass}^{I=1/2,1}$ is imported from Eq. (\ref{L_mass_I1}) together with 
its parameters from Eq. 
(\ref{a0k0_m_bare}).

The mixing between $N$ and 
$N'$, and the mixing of these two nonets with the  scalar glueball $G$ can 
be 
written as
\begin{equation}
{\cal L}_{mix}^{I=0} = 
{\cal L}_{mix}^{I=1/2,1}
- \rho {\rm Tr} (N) {\rm Tr} (N')
- e G {\rm Tr} \left( N \right)
- f G {\rm Tr} \left( N' \right)
\label{L_mix_I0}
\end{equation}
where the first term is given in 
(\ref{L_mix_I1}) with 
$\gamma$ from (\ref{a0k0_m_bare}).   The second term does not 
contribute to the $I=1/2,1$ mixing, and in  special limit of 
$\rho\to-\gamma$:
\begin{equation}
- \gamma  {\rm Tr} (N N')
- \rho {\rm Tr} (N) {\rm Tr} (N')
= \gamma \epsilon^{abc} \epsilon_{ade} N^d_b {N'}^e_c
\label{mix_OZI}
\end{equation} 
which is more consistent with the OZI rule than 
the individual $\gamma$ and $\rho$ terms and is studied in 
\cite{02_TKM}.    
Here we do not restrict the mixing to this particular combination, and 
instead, take $\rho$ as a a priori unknown free parameter.
Terms with unknown couplings $e$ and $f$
describe mixing with the scalar glueball $G$.    
As a result, the five isosinglets 
below
2 GeV, become a mixture of five different flavor combinations, and their 
masses can be organized as
\begin{equation}
{\cal L}_{mass}^{I=0} + {\cal L}_{mix}^{I=0} = - {1\over 2} {\tilde {\bf 
F}}_0 {\bf M}^2 {\bf 
F}_0 = 
- {1\over 2} {\tilde {\bf F}} {\bf M}_{diag.}^2 {\bf F}  
\label{L_mass_mix}
\end{equation}
with
\begin{equation}
{\bf F}_0 =
       \begin{array}{l}
                         \left(
                         \begin{array}{c}
                              N_3^3 \\
                             (N_1^1 + N_2^2)/\sqrt{2}\\
                              {N'}_3^3 \\
                             ({N'}_1^1 + {N'}_2^2)/\sqrt{2}\\
                              G  
                         \end{array}
                         \right)
                   
       \end{array}   
=
       \begin{array}{l}
                         \left(
                         \begin{array}{c}
      {\bar u}{\bar d} u d\\
     ({\bar s}{\bar d} d s + {\bar s}{\bar u} u s) /\sqrt{2}\\
                              {\bar s}s\\
                             ({\bar u} u + {\bar d} d)/\sqrt{2}\\
                              G  
                         \end{array}
                         \right)
                   
       \end{array}   
=
       \begin{array}{l}
                         \left(
                         \begin{array}{l}
                             f_0^{NS}\\
                             f_0^S \\
                             {f'}_0^S\\
                             {f'}_0^{NS} \\
                              G  
                         \end{array}
                         \right)
       \end{array}   
\label{SNS_def}
\end{equation}
where the superscript $NS$ and $S$ respectively represent the non-strange 
and strange combinations.  ${\bf F}$ contains the physical fields
\begin{equation}
{\bf F}  =
       \begin{array}{c}
                        \left( \begin{array}{c}
                                \sigma(550)\\
                                f_0(980)\\
                                f_0(1370)\\
                                f_0(1500)\\
                                f_0(1710)
                                \end{array}
                         \right)  
= K^{-1} {\bf F}_0
       \end{array}   
\label{K_def}
\end{equation}
where $K^{-1}$ is the transformation matrix.
The mass squared matrix is
\begin{equation}
{\bf M}^2  =
\left[ 
\begin{array}{ccccc}
2 m_{K_0}^2 - m_{a_0}^2  + 2 (c + d x) &
\sqrt{2}[2c + (1+x)d] &
\gamma + \rho &
\sqrt{2} \rho &
e
\\
\sqrt{2}[2c + (1+x)d] &
m_{a_0}^2  + 4 (c + d) &
\sqrt{2} \rho &
\gamma + 2 \rho&
\sqrt{2} e 
\\
\gamma + \rho&
\sqrt{2} \rho &
2 m_{K'_0}^2 - m_{a'_0}^2  + 2 (c' + d' x) &
\sqrt{2}[2c' + (1+x)d'] &
f
\\
\sqrt{2} \rho &
\gamma + 2 \rho&
\sqrt{2}[2c' + (1+x)d'] &
m_{a'_0}^2 + 4 (c' + d') &
\sqrt{2} f
\\
e &
\sqrt{2} e &
f &
\sqrt{2} f &
2 g
\end{array}
\right]
\label{mass_matrix}
\end{equation} 
in which the value of the unmixed $I=1/2,1$ masses, and
the mixing parameter $\gamma$ are substituted in from
(\ref{bare_masses}).   We see that there are eight unknown parameters in
(\ref{mass_matrix}) which are $c$, $d$, $c'$, $d'$, $g$, $\rho$, $e$ and 
$f$.        We will use numerical analysis to search this eight 
dimensional parameter space for the best values that give closest 
agreement with experimentally known masses. Particularly, we will  
study in detail 
the effect of the large uncertainties on the mass of the $f_0(600)$ and 
$f_0(1370)$ on the resulting parameters.


\section{Matching the Theoretical Prediction to Experimental Data}

\begin{figure}[h]
\epsfxsize=6cm
\epsfbox{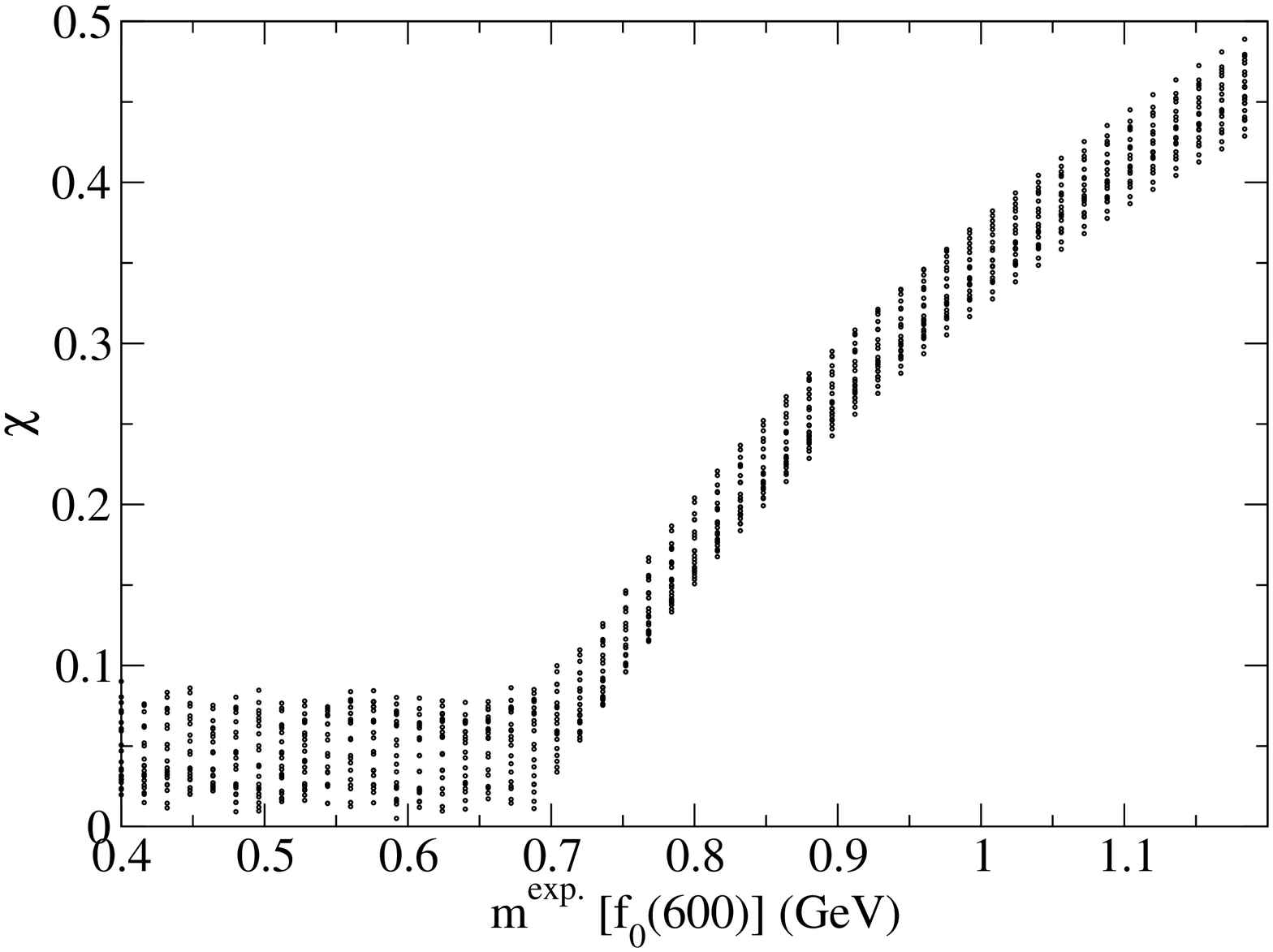}
\hskip .5cm
\epsfxsize=6cm
\epsfbox{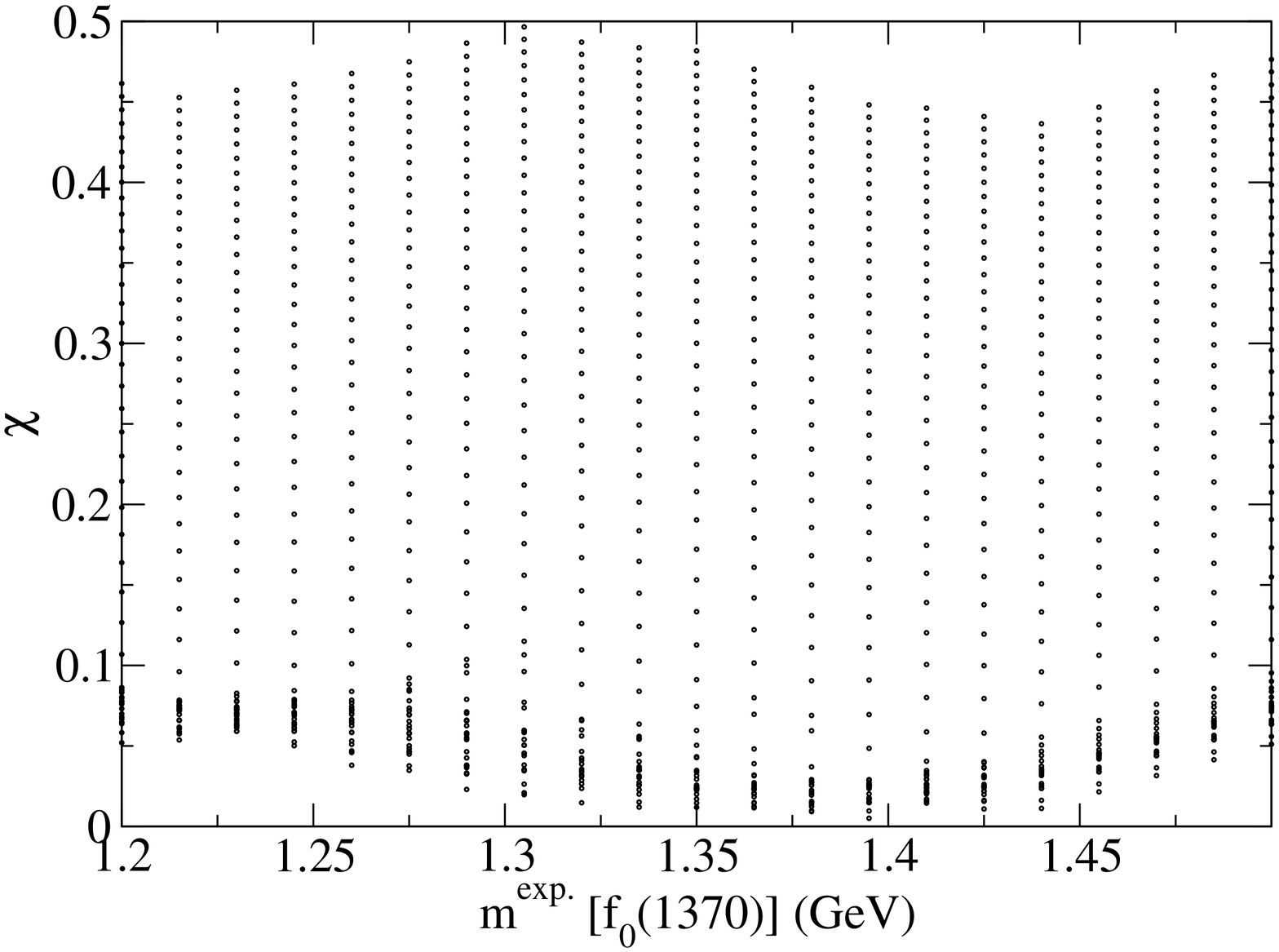}
\vspace*{8pt}
\caption{
	Projection of $\chi$ onto the $\chi-m^{\rm exp.} 
[f_0(600)]$ plane, and onto the $\chi-m^{\rm exp.} 
[f_0(1370)]$ plane.  The left figure shows
that values of $m^{\rm exp.} [f_0(600)] >$ 700 MeV result
in a significant disagreement between theory and
experiment.   The right figure shows that $m^{\rm exp.} 
[f_0(1370)]$  outside of the range 1300-1450 MeV is not 
favored by the model.
	\label{F_chi_vs_msf}}
\vskip .5cm
\end{figure}

To determine the eight unknown Lagrangian parameters 
($c,c',d,d',e,f,g$ and
$\rho$), we input the
experimental masses of the scalar states.  Out of the five
isosignlet states, three have a well established
experimental mass \cite{PDG}: 
\begin{equation}
\begin{array}{r c l} 
m^{\rm exp.} [f_0(980)] &=& 980 \pm 10 \, {\rm MeV} \\ 
m^{\rm exp.} [f_0(1500)] &=& 1507 \pm 5 \, {\rm MeV} \\ 
m^{\rm exp.} [f_0(1710)] &=& 1713 \pm 6 \, {\rm MeV}
\label{exp_masses} 
\end{array} 
\end{equation} 
However, the experimental mass of the $f_0(600)$
and the $f_0(1370)$ have very large uncertainties
\cite{PDG}:  
 \begin{eqnarray} 
m^{\rm exp.} [f_0(600)] &=& 400
\rightarrow 1200 \,\,{\rm MeV} \nonumber \\ 
m^{\rm exp.} [f_0(1370)] &=& 1200 \rightarrow 1500 \,\,{\rm 
MeV}
\label{13_masses} 
\end{eqnarray}
We search for the eight Lagrangian parameters 
(which
in turn determine the mass matrix (\ref{mass_matrix}))  
such that three of the resulting eigenvalues match their
central experimental values in (\ref{exp_masses}), and
the other two masses fall somewhere in the
experimental ranges in (\ref{13_masses}).  
This means that out of the five eigenvalues of (\ref{mass_matrix}), 
three (eigenvalues 2, 4 and 5) should match the fixed target values 
in (\ref{exp_masses}),  but two (eigenvalues 1 and 3) have no fixed target 
values and instead can 
be match to any values in (\ref{13_masses}).
Therefore, to include all possibilities for the experimental ranges 
in (\ref{13_masses}), we numerically
scan the $m^{\rm exp.} [f_0(600)]$$-$$m^{\rm exp.} [f_0(1370)]$
plane over the allowed ranges,  
and at each point we fit for the eight Lagrangian 
parameters
such that the three theoretically calculated eigenvalues (2, 4 and 5) 
match
the fixed target masses in (\ref{exp_masses}),  and the
other two eigenvalues (1 and 3) match the variable target masses at the  
chosen point in this plane.  At a given point in the 
$m^{\rm exp.} [f_0(600)]$$-$$m^{\rm exp.} [f_0(1370)]$ plane, 
we measure
the goodness of the fit by the smallness of the quantity
        \begin{equation} 
\chi \left(
m^{\rm exp.}[f_0(600)], m^{\rm exp.}[f_0(1370)]\right) = \sum_i {
  {|m^{\rm theo.}_i - m^{\rm exp.}_i|}  
             \over
      m^{\rm exp.}_i
}
\end{equation} 
where $i=1 \cdots 5$ correspond to the five $I=0$
states in ascending order (for example $m^{\rm exp.}_1$ =
$m^{\rm exp.} [f_0(600)]$, $\cdots$). 
The value of $\chi\times 100$ gives 
the overall percent difference between theory and
experiment. 
The total of 14,641 eight-parameter fits were 
performed over the target points in the $m^{\rm exp.} 
[f_0(600)]$$-$$m^{\rm exp.} 
[f_0(1370)]$ plane
(the overall completion of the numerical analysis of this article 
required several months of computation time on a XEON dual-processor 
workstation).  
This
procedure creates a three dimensional graph of $\chi$ as a
function of $m^{\rm exp.}[f_0(600)]$ and
$m^{\rm exp.}[f_0(1370)]$. The overall result of the fits for
the experimentally allowed range of masses are given in
Fig. \ref{F_chi_vs_msf}, in which the projection of
$\chi$ onto the $\chi-m^{\rm exp.}[f_0(600)]$ plane and onto
the $\chi-m^{\rm exp} [f_0(1370)]$ plane are shown.  We can
easily see that the experimental mass of $f_0(600)$ above
700 MeV, and the experimental mass of $f_0(1370)$ outside
the range of 1300 to 1450 MeV are not favored by this
model.   

\begin{figure}[h]
\epsfxsize=6cm
\epsfbox{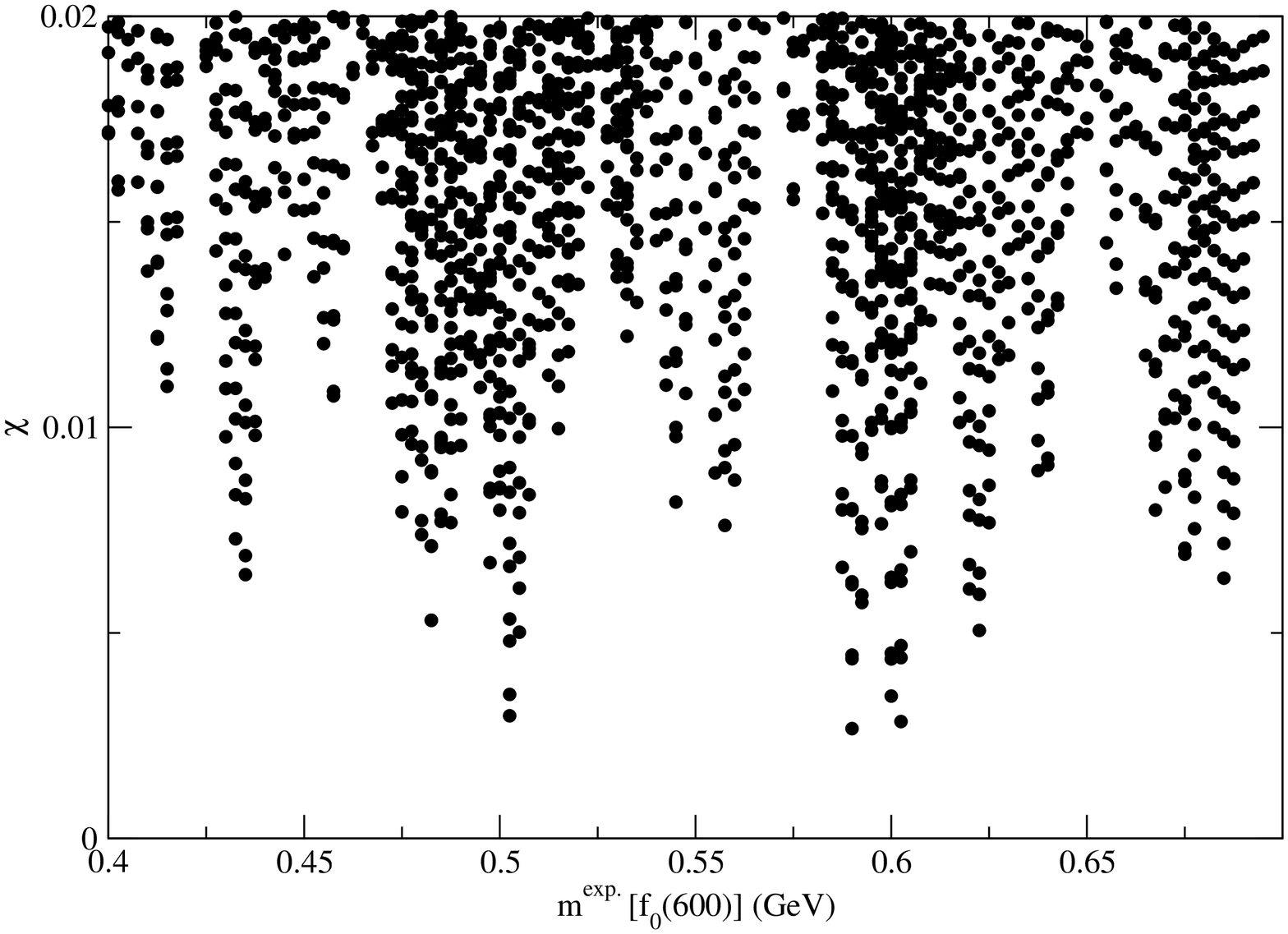}
\hskip .5cm
\epsfxsize=6cm
\epsfbox{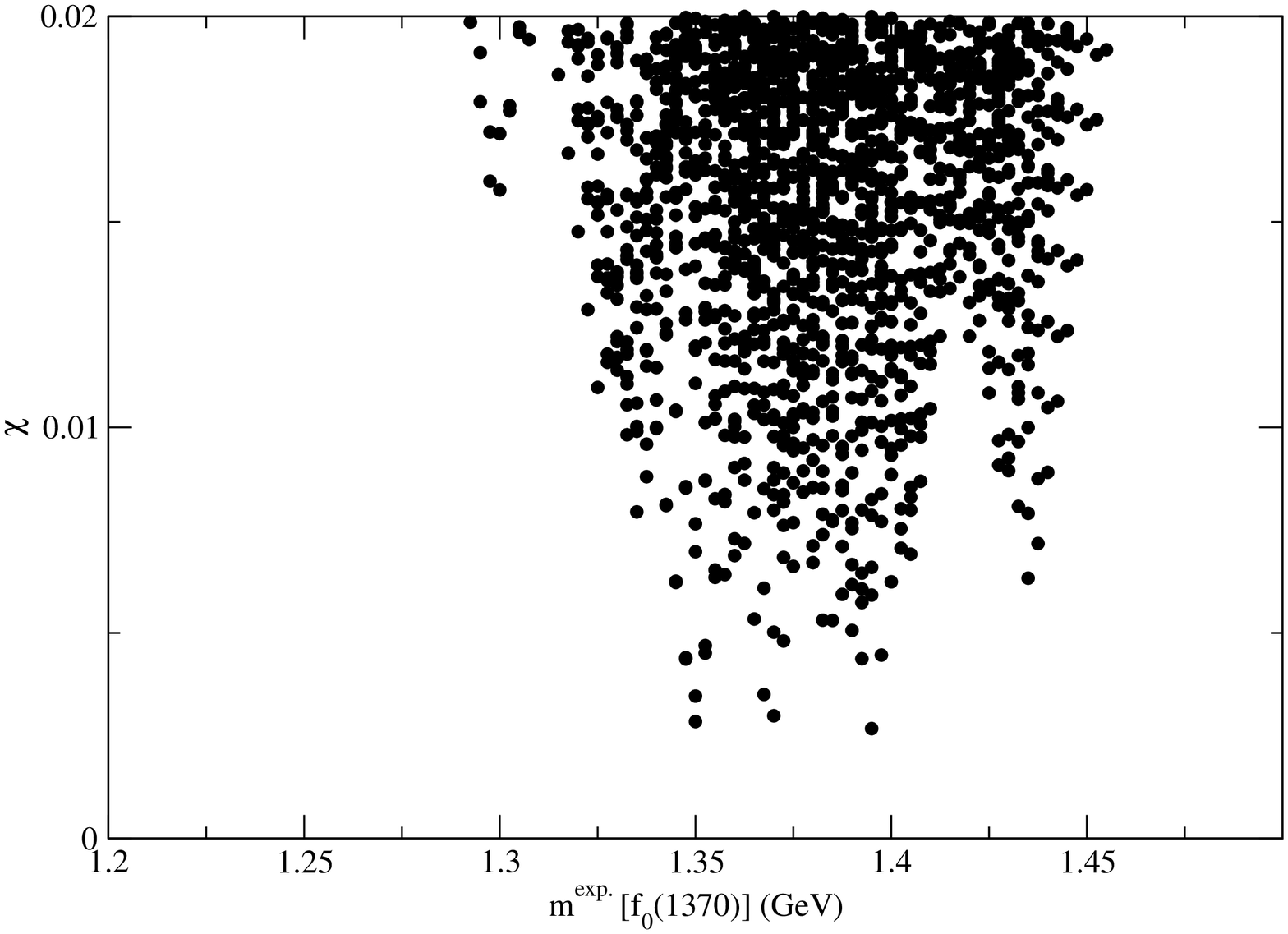}
\vspace*{8pt} 
\caption{ 
	Projection of $\chi$ onto the $\chi-m^{\rm exp.} 
[f_0(600)]$ plane, and onto the $\chi-m^{\rm exp.} 
[f_0(1370)]$ plane.  The left figure shows
a series of local minima in $\chi$ with the lowest values
in the range 500 MeV $\le m^{\rm exp.} [f_0(600)] \le$
600 MeV.  The right figure shows that the present model
predicts that $m^{\rm exp.} [f_0(1370)]$ is confined
within a smaller range of 1300 MeV to 1450 MeV.  \\
	\label{F_chi_vs_msf_2}} 
\end{figure}

\begin{figure}[h]
\epsfxsize=6cm
\epsfbox{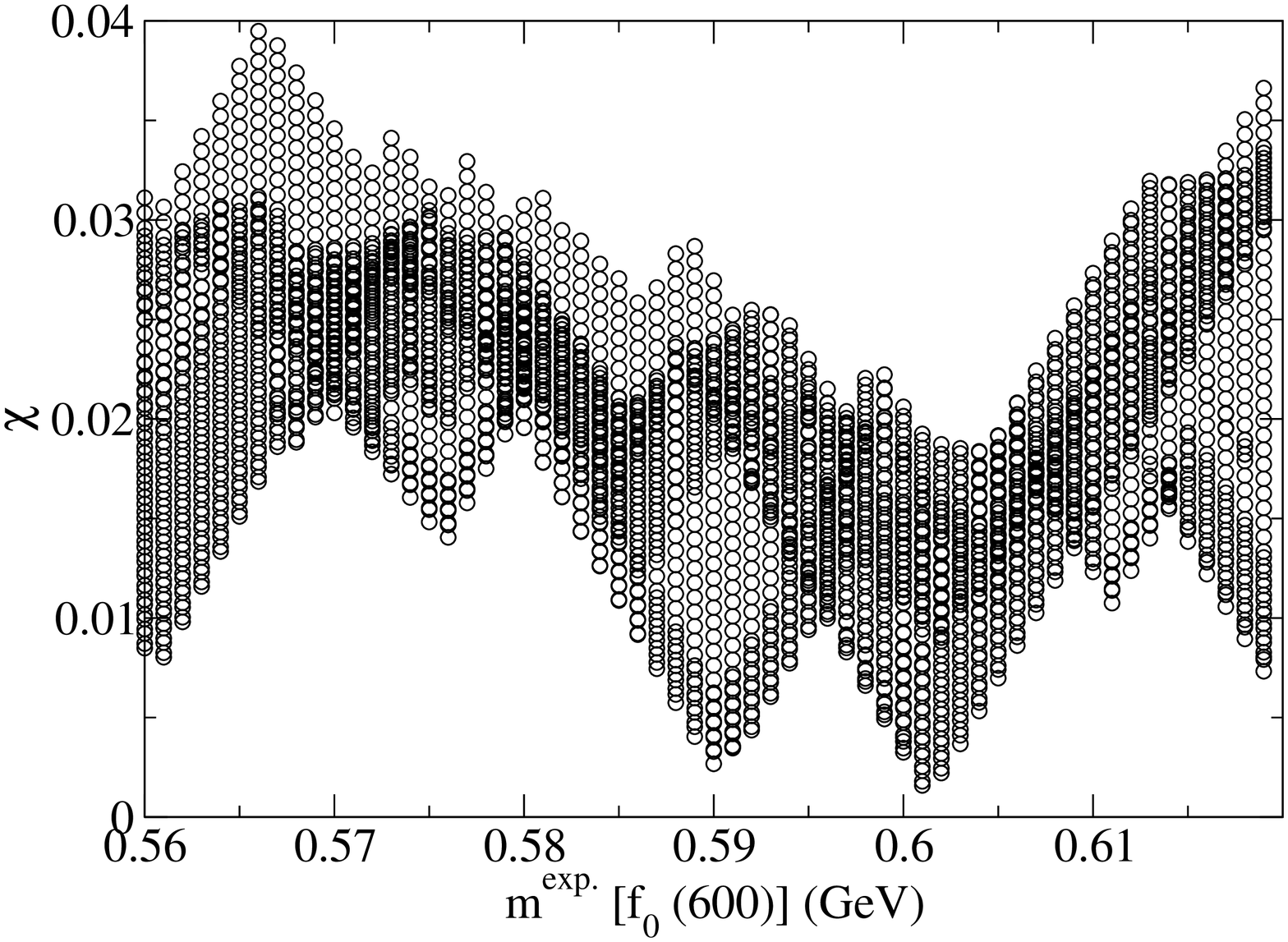}
\hskip .5cm
\epsfxsize=6cm
\epsfbox{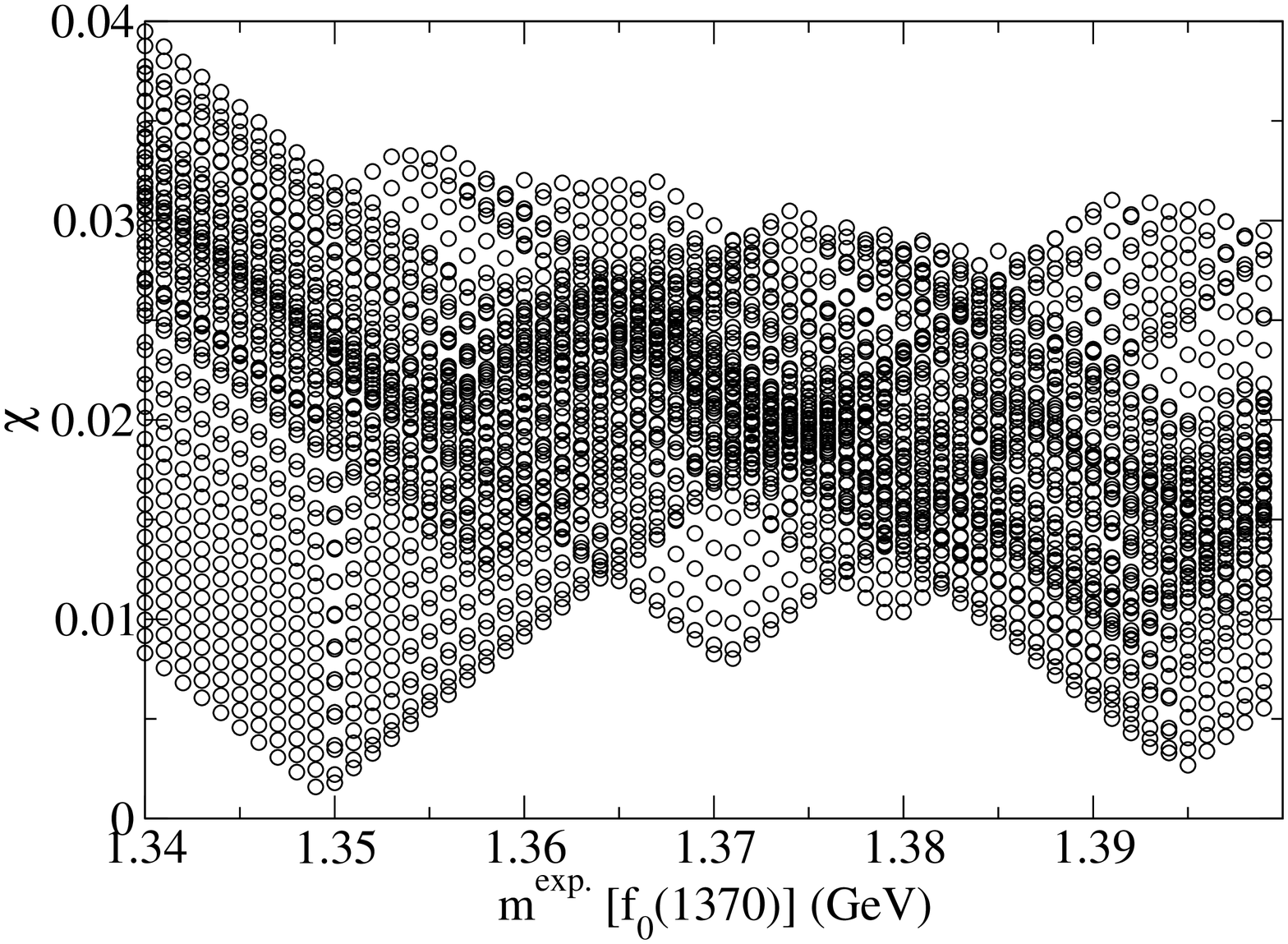} 
\vspace*{8pt} 
\caption{Projection of $\chi$ onto the 
$\chi-m^{\rm exp.} [f_0(600)]$ plane, and onto  
the $\chi-m^{\rm exp.} 
[f_0(1370)]$ plane.   $\chi$ exhibits a series of local 
minima.\\
\label{F_chi_best}} 
\end{figure}

\begin{figure}[h]
\epsfxsize=6cm
\epsfbox{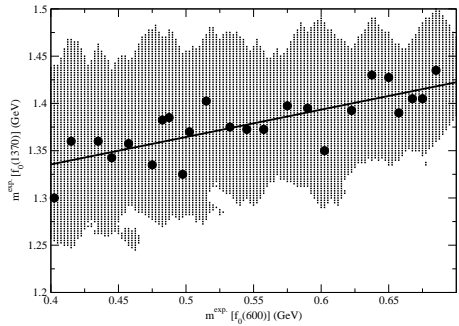}
\vspace*{8pt}
\caption{Projection of $\chi$ onto the $m^{\rm exp.} 
[f_0(600)]$$-$$m^{\rm exp.}[f_0(1370)]$ plane.  In the gray 
region there is an overall disagreement of less than 5\% 
between theory and experiment. The circles 
represent points on this plane at which $\chi$ has a 
local minimum and $\chi < 0.02$.   The solid line 
captures the trend of these local minima with  
equation: 
$m [f_0(1370)] = 0.29 \, m 
[f_0(600)] + 1.22 $ GeV.
\label{F_ms_mf_plane}} 
\end{figure}

\begin{figure}[h]
\epsfxsize=4cm
\epsfbox{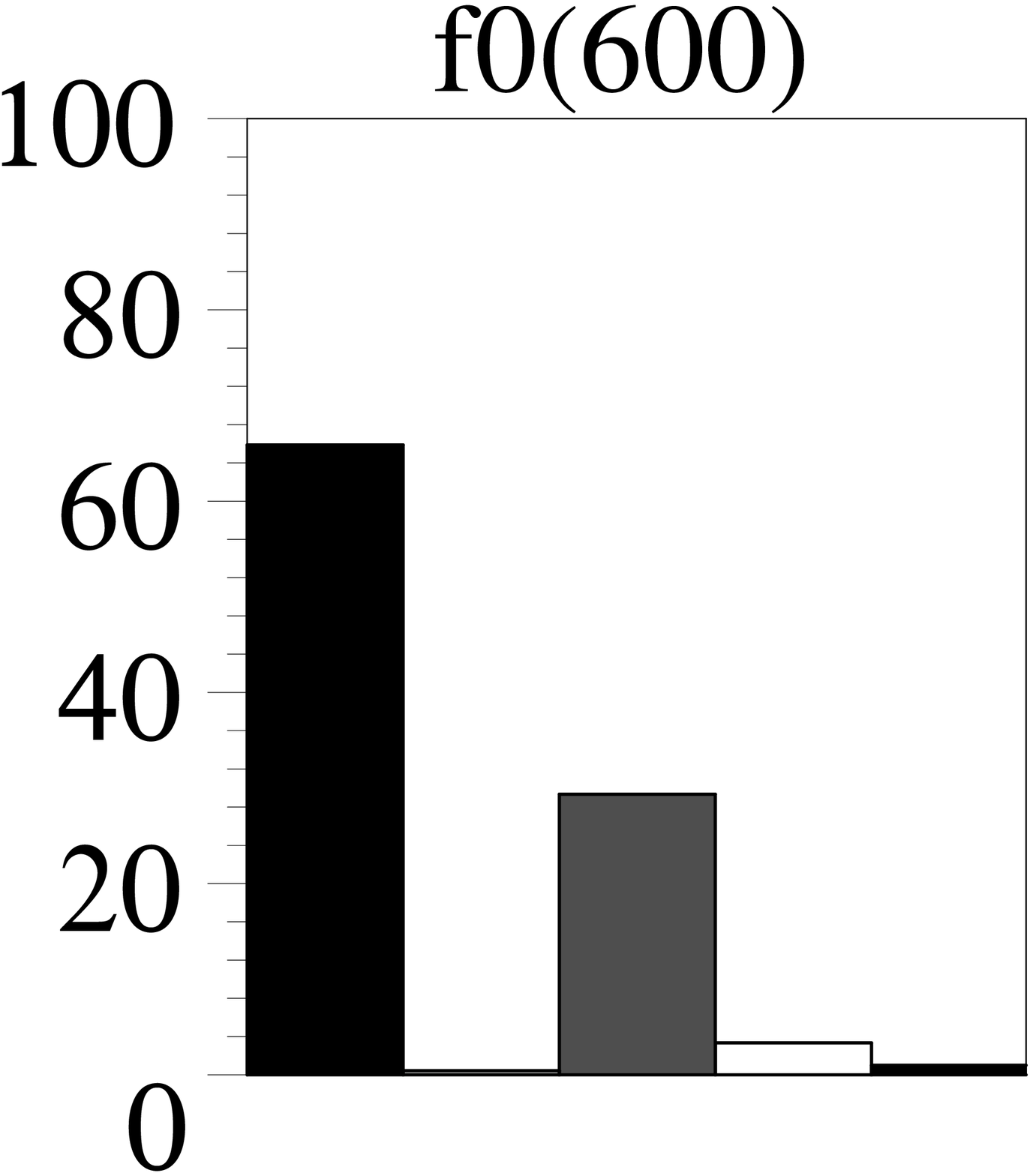}
\epsfxsize=4cm
\epsfbox{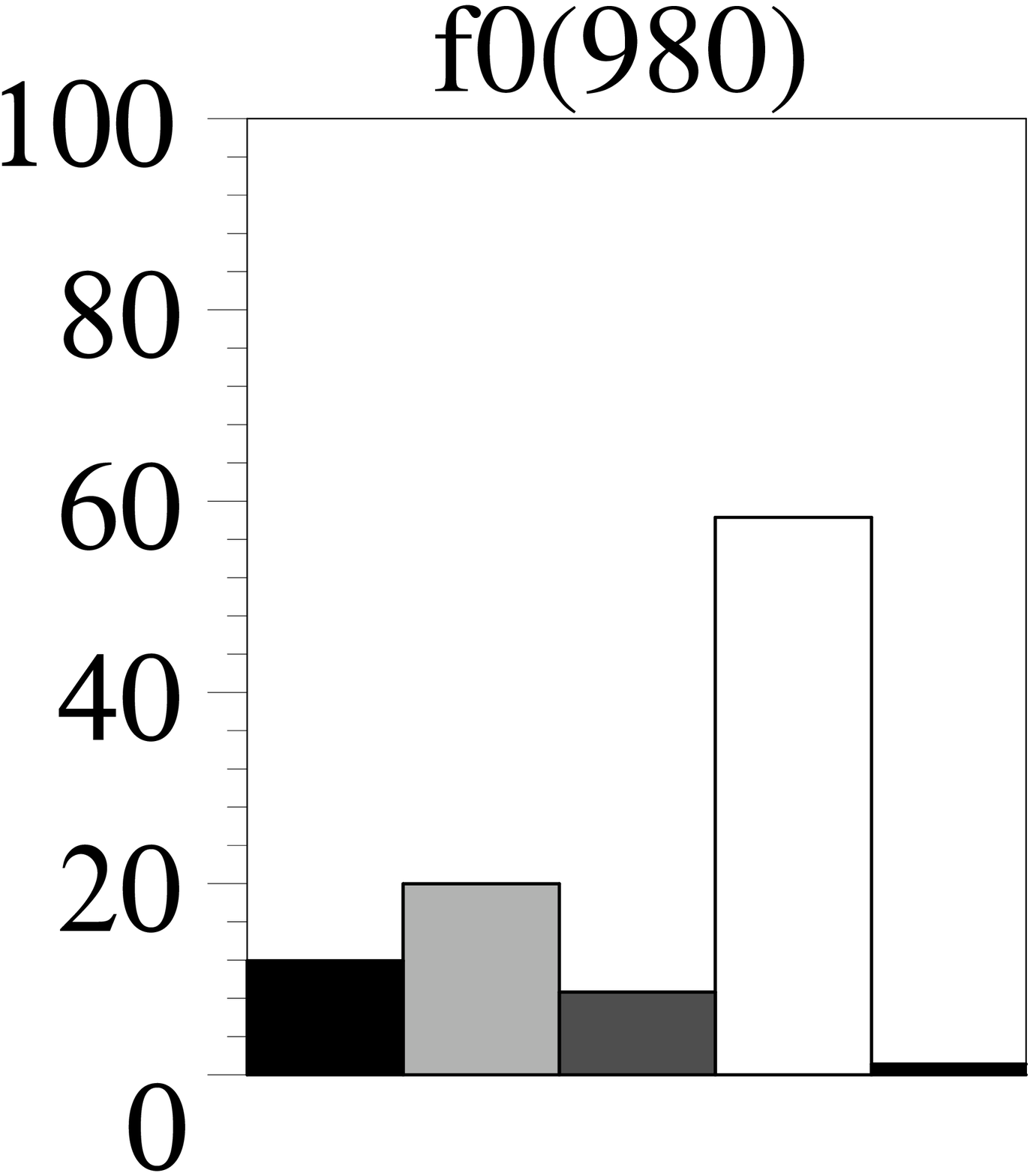}
\epsfxsize=4cm
\epsfbox{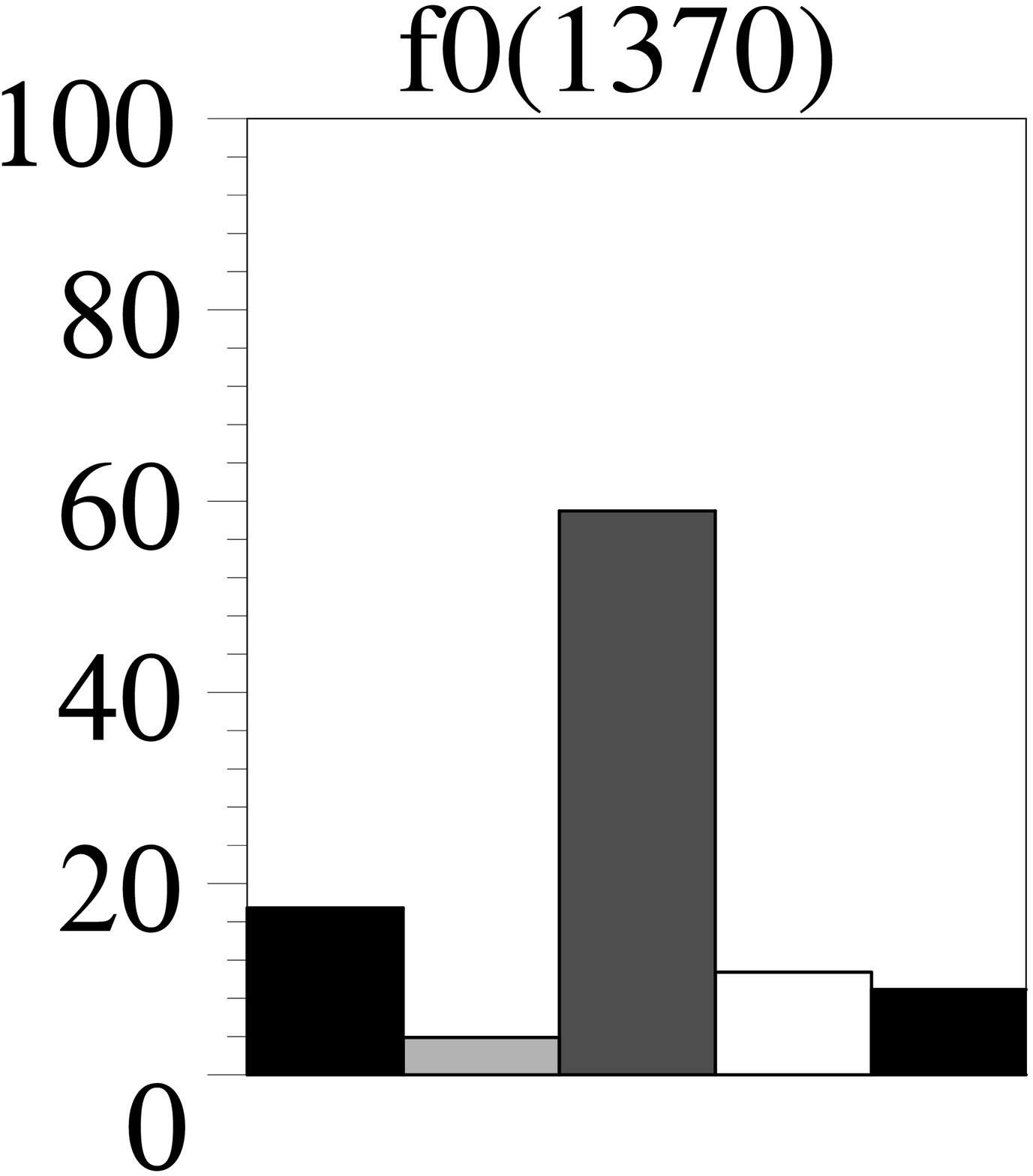}
\epsfxsize=4cm
\epsfbox{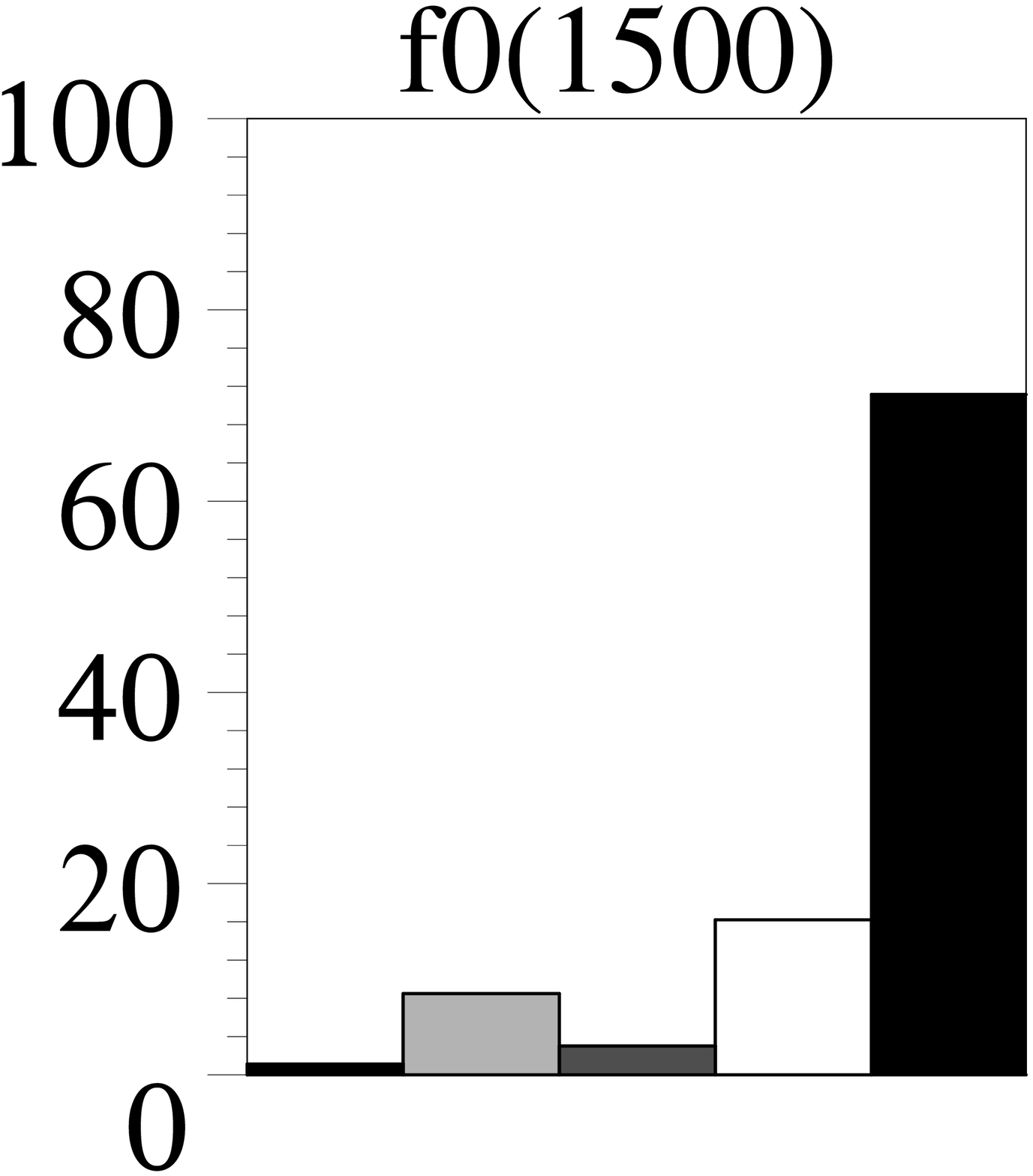}
\epsfxsize=4cm
\epsfbox{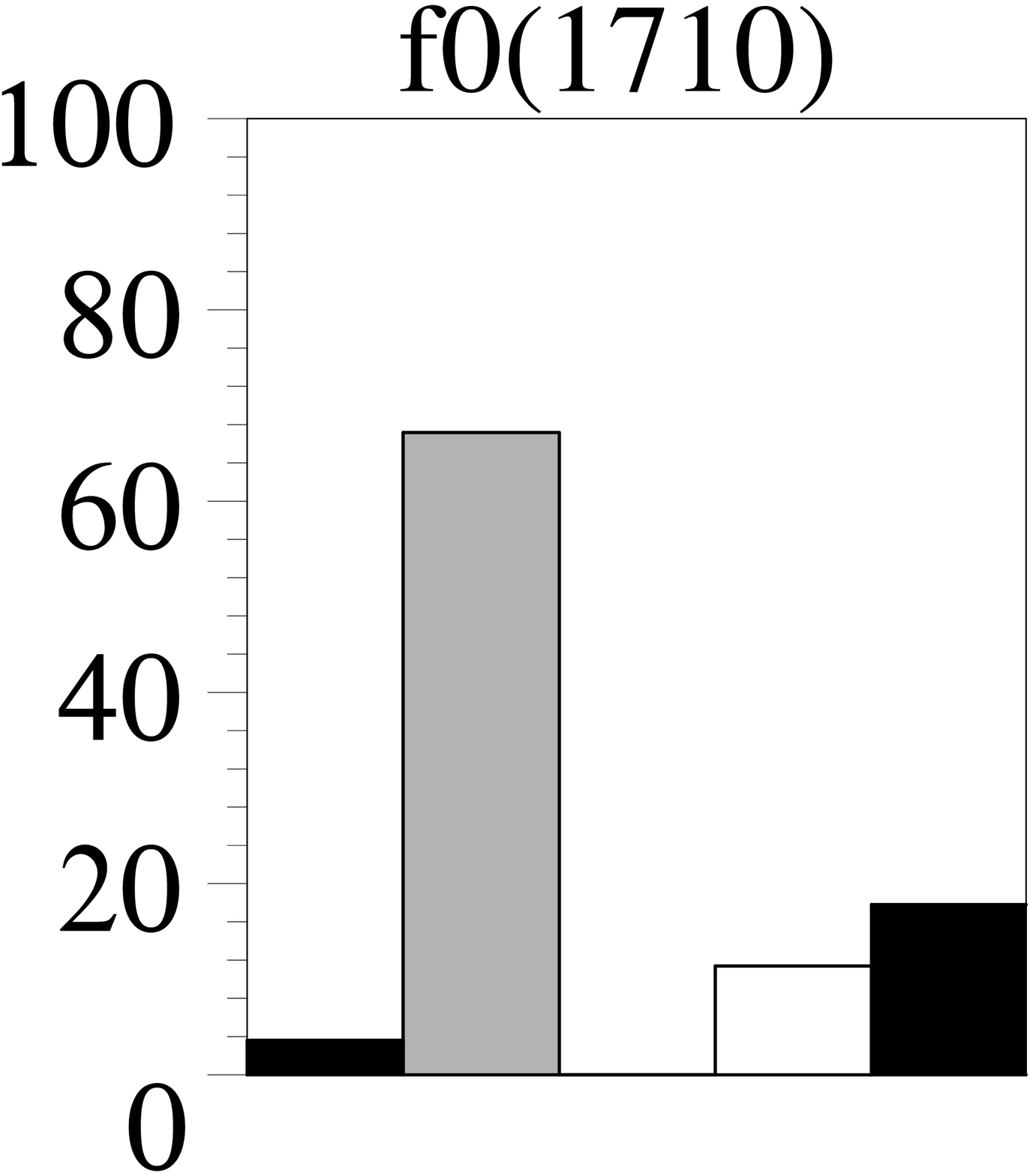}
\vspace*{8pt}
	\caption{Percentage of the quark and glueball
components of the scalar states for values of 
$m[f_0(600)]$ = 558 MeV and $m [f_0(1370)]$ = 1373  MeV. 
In each 
figure, the
columns from left to right respectively represent the
percentage of:  ${\bar u}{\bar d} u d$ (black),
$({\bar s}{\bar d} d s + {\bar s}{\bar u} u s) /\sqrt{2}$
(light gray), ${\bar s}s$ (dark gray), $({\bar u} u + 
{\bar
d} d)/\sqrt{2}$ (white), and glueball (black).
	\label{F_comps}} 
\end{figure}

More refined numerical analysis narrows down the favored
regions to the ranges shown in Fig. \ref{F_chi_vs_msf_2}.  
We see that the lowest value of $\chi$ occurs within the
ranges: 
\begin{eqnarray} 
m^{\rm exp.} [f_0(600)] &=& 500\rightarrow 600 \, 
{\rm MeV}\nonumber 
\\ m^{\rm exp.}
[f_0(1370)] &=& 1350 \rightarrow 1400\, {\rm MeV}
\label{chi_min} 
\end{eqnarray} 
	In this region, $\chi$ exhibits a series of local
minima, also shown in a more refined numerical work in
Fig. \ref{F_chi_best}. Although $\chi$ has its lowest
values in the regions given in (\ref{chi_min}) we see
that it only increases by approximately 0.01 outside of
this region and can be comparable to the theoretical
uncertainties of this framework.  Therefore, for a more
conservative estimate we take into account all local
minima with $\chi <0.02$.  This confines the experimental
masses to the ranges:
	\begin{eqnarray}
m^{\rm exp.} [f_0(600)] &=& 400 \rightarrow 700 \, {\rm MeV}\nonumber \\
m^{\rm exp.} [f_0(1370)] &=& 1300 \rightarrow 1450\, {\rm  MeV}
\label{chi_min2}
\end{eqnarray}
	The local minima of $\chi$ in this range are
shown in Fig. \ref{F_ms_mf_plane} in which the projection
of $\chi$ onto the $m^{\rm exp.} [f_0(600)]$$-$$m^{\rm exp.}
[f_0(1370)]$ plane is given.  In the gray region, there
is an overall disagreement of less than 5\% between
theory and experiment.  The local minima (with $\chi
<0.02$) are shown with dots, together with a linear fit
that shows the correlation between the mass
of the $f_0(600)$ and the $f_0(1370)$:  
\begin{equation}
m [f_0(1370)] = 0.29 \, m [f_0(600)] + 1.22\,\,{\rm   GeV}
\label{linearfit}
\end{equation}

	At all local minima (with $\chi <0.02$),
a detailed numerical analysis is performed and the eight
Lagrangian parameters are determined.   The large 
uncertainties on $m[f_0(600)]$ and $m[f_0(1370)]$ do not 
allow an accurate determination of these parameters.    
In this work we study the correlation between these 
uncertainties and determination of the Lagrangian 
parameters.  For orientation, let us first begin by 
investigating  
the central
values of 
\begin{eqnarray} 
m^{\rm exp.} [f_0(600)] &=& 558 \,\,{\rm MeV} \nonumber 
\\ 
m^{\rm exp.} [f_0(1370)] &=& 1373 \,\,{\rm MeV} 
\label{m13_central} 
\end{eqnarray}
 	It is important to also notice that the central value
of $m[f_0(600)]$ = 558 MeV is exactly what was first found
in \cite{San} in applying the nonlinear chiral Lagrangian of this work to 
$\pi\pi$
scattering. 
	At the particular point of (\ref{m13_central}), the result of the
fit is given in Table \ref{T_fit_central}, and is consistent with the
initial investigation of this model in ref. \cite{04_F1} in which the 
effect
of the mixing parameter $\rho$ is only studied at several discrete points.  
The result given in Table \ref{T_fit_central} supplements the work of
\cite{04_F1} by treating $\rho$ as a general free parameter. The rotation
matrix is
\begin{equation}
K^{-1} =
\left[
\begin{array}{ccccc}
  0.812 & -0.065 & -0.542 & -0.183 &  0.101 
\\
   0.346  & -0.447  &  0.294  & 0.764  & -0.106 
\\
   0.418  &   0.198   &  0.768  & -0.328   &  0.299 
\\
  -0.106  &   0.292  &  -0.174   &   0.403   & 0.844 
\\
 0.190  & 0.820  &  -0.006  & 0.338   & -0.422 
\end{array}
\right]
\label{K_inverse_central}
\end{equation}
and in turn
determines the quark and glueball components of each physical state as
presented in Fig. \ref{F_comps}.  Several general observations can be made
in Fig.  \ref{F_comps}.  Clearly, the $f_0(600)$ is dominantly a
non-strange four-quark state with a substantial ${\bar s} s$ component.  
On the other hand the dominant structure of the $f_0(1370)$ seems to be
the reverse of the $f_0(600)$.  The $f_0(980)$ seems to have a dominant
non-strange two-quark component with a significant strange four-quark
component. The $f_0(1500)$ appears to have a dominant glueball component
with some minor two and four quark admixtures. The dominant component of
the $f_0(1710)$ seems to be a strange four-quark state followed by a
glueball component.

\begin{table}[h]
\begin{center}
\begin{tabular}{c||c}
\hline
\hline
Lagrangian Parameters & Fitted Values (GeV$^2$)
\\
\hline 
\hline 
$c$  & 2.32 $\times 10^{-1}$ 
 \\
$d$  & $-9.11 \times 10^{-3}$ 
 \\
$c'$ & $-4.29 \times 10^{-3}$ 
  \\
$d'$ &$-1.25 \times 10^{-2}$
\\
$g$ & 1.16  
 \\
$\rho$ &$2.87 \times 10^{-2}$ 
\\
$e$ &$-2.13\times 10^{-1}$ 
\\
$f$ &$6.10 \times 10^{-2}$ 
\\
\hline
\end{tabular}
	\caption{Fitted values of the Lagrangian
parameters for $m[f_0(600)]$ = 558 MeV and $m 
[f_0(1370)]$ = 1373  MeV.} 
\label{T_fit_central} 
\end{center} 
\end{table}

The scalar glueball mass can be calculated from the fit 
in Table \ref{T_fit_central}:
\begin{equation}
m_G = \sqrt{2\,\, g} = 1.52  \,\, {\rm GeV} \\
\end{equation}
which is in a range expected from Lattice QCD \cite{Lattice}.

Next we will examine
the effect of deviation from the central values of
$m[f_0(600)]$ and $m[f_0(1370)]$ on the predictions given
in Fig.  \ref{F_comps}.  We will find that the average
properties of the $f_0(600)$, $f_0(980)$ and $f_0(1370)$
remain close to those given in Fig. \ref{F_comps},  but some of  the 
properties of the $f_0(1500)$ and $f_0(1710)$ are 
sensitive to such deviations.
 
To investigate the effect of the mass uncertainties of 
the $f_0(600)$ and $f_0(1370)$, we perform 8-parameter 
fits at each of the local minima of Fig. 
\ref{F_ms_mf_plane}, and determine the variation of the 
Lagrangian parameters around the values in Table 
\ref{T_fit_central}.   The results are summarized in 
Table \ref{T_fit_average}.    We see that practically the Lagrangian 
parameters are quite sensitive to the mass of the $f_0(600)$ and 
$f_0(1370)$.    The only exception is parameter $g$ which determines the 
glueball mass:
\begin{equation}
m_G = \sqrt{2\,\, g} = 1.5\rightarrow 1.7 \,\, {\rm GeV} 
\label{Gmass_range}
\end{equation}

\begin{table}[h]
\begin{center}
\begin{tabular}{c||c|c}
\hline
\hline
Lagrangian Parameters & Average (GeV$^2$)  & Variation (GeV$^2$)
\\
\hline 
\hline 
$c$  & $1.71  \times 10^{-1}$
& ($0.37 \rightarrow 2.60 )  \times 10^{-1}$ 
 \\
$d$  & $ - 7.32 \times 10^{-3}$
& ($-14.44 \rightarrow -2.63) \times 10^{-3}$ 
 \\
$c'$ & $ - 2.87  \times 10^{-3}$
& ($-5.32 \rightarrow -0.64) \times 10^{-3}$ 
  \\
$d'$ & $ -0.97  \times 10^{-2}$
& ($-1.41 \rightarrow -0.27) \times 10^{-2}$
\\
$g$ & 1.24 
& 1.15 $\rightarrow$ 1.44  
 \\
$\rho$ & $4.7  \times 10^{-2}$
& ($1.31 \rightarrow 9.14) \times 10^{-2}$ 
\\
$e$ & $- 2.10  \times 10^{-1}$
& ($-3.02 \rightarrow -1.76) \times 10^{-1}$ 
\\
$f$ & $ 3.74 \times 10^{-2}$
& ($1.15 \rightarrow 11.40) \times 10^{-2}$ 
\\
\hline
\end{tabular}
	\caption{ 
	The effect of the uncertainties of $m^{\rm exp.} [f_0(600)]$ and 
$m^{\rm exp.}[f_0(1370)]$ on the Lagrangian parameters.  The averaged value of the
parameters (second column) is compared with their (asymmetric) range of
variation (third column).  Only parameter $g$ (which determines the
glueball mass) is relatively insensitive to $m^{\rm exp.}[f_0(600)]$ and 
$m^{\rm exp.}[f_0(1370)]$. } 
	\label{T_fit_average} 
\end{center} 
\end{table}

\begin{figure}[h]
\epsfxsize=5cm
\epsfbox{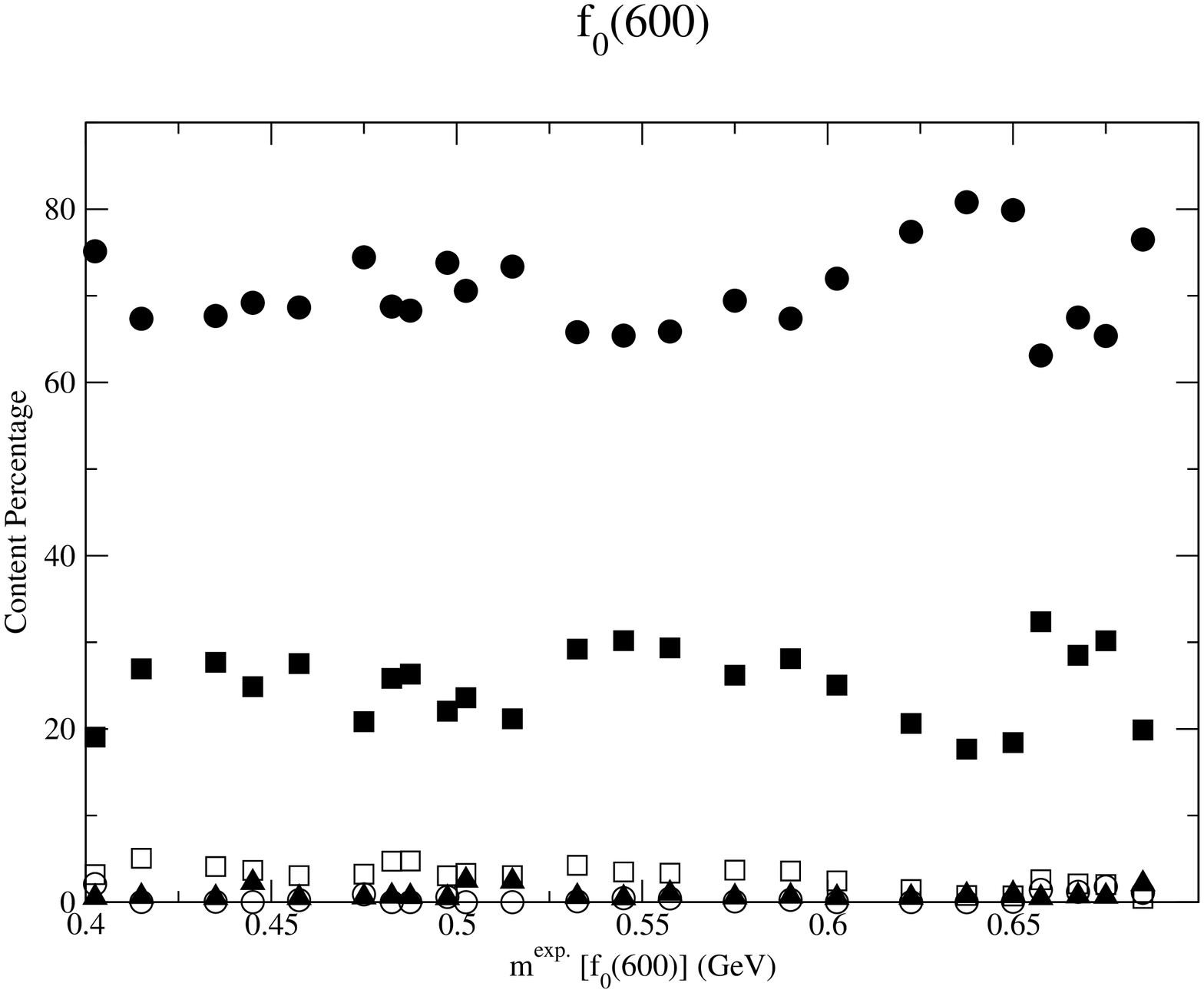}
\hskip .5cm
\epsfxsize=5cm
\epsfbox{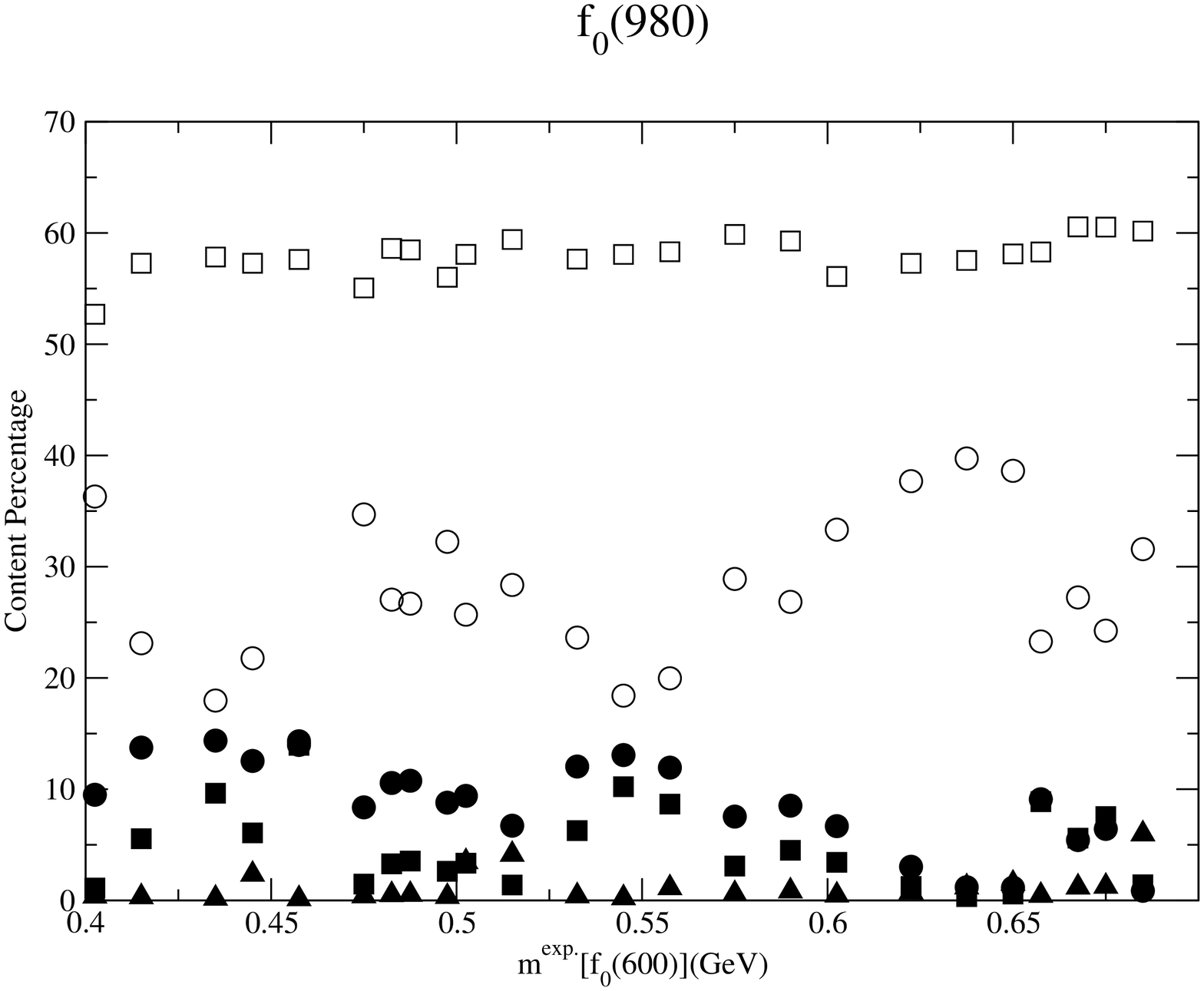}
\vskip .5cm
\epsfxsize=5cm
\epsfbox{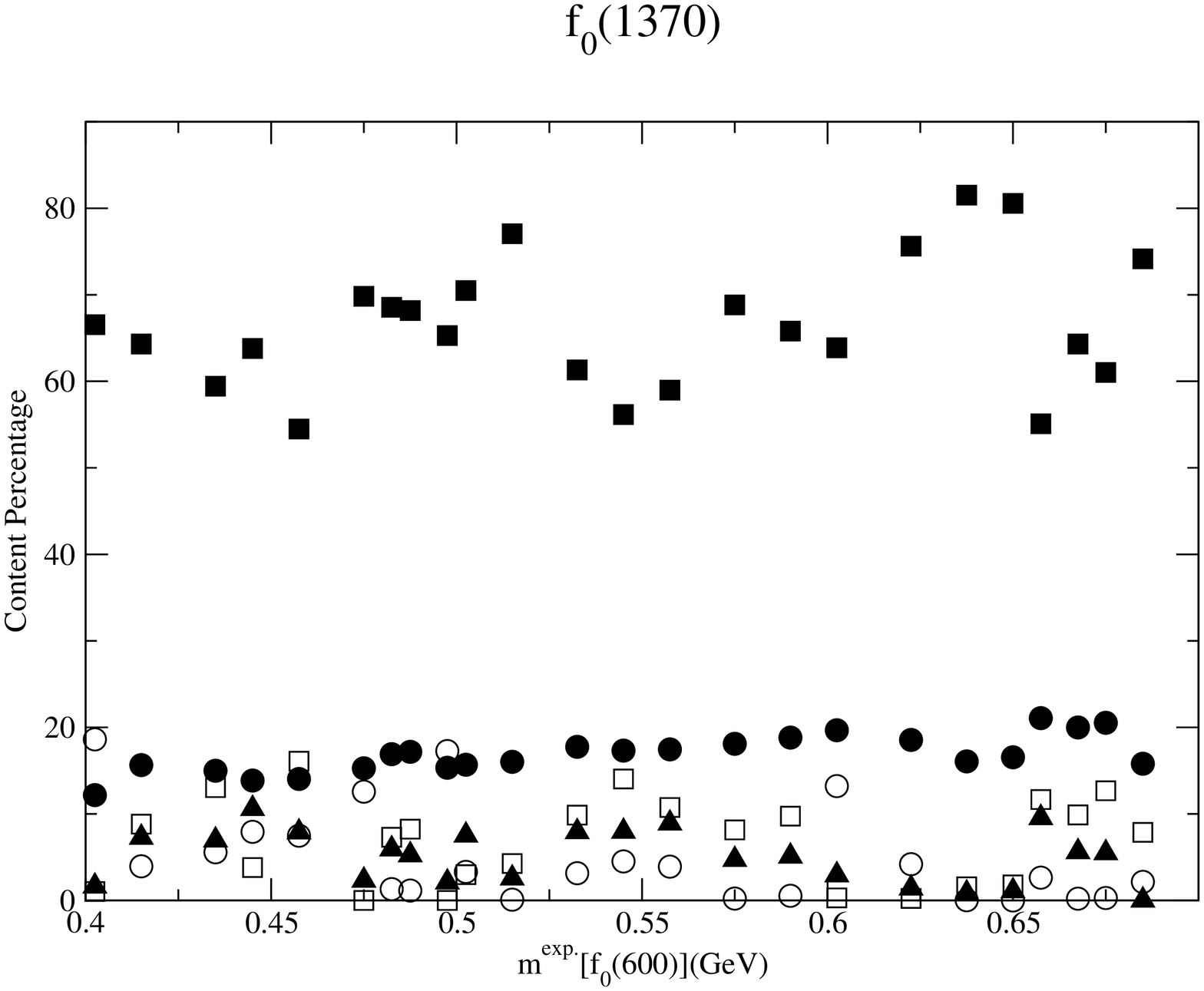}
\hskip .5cm
\epsfxsize=5cm
\epsfbox{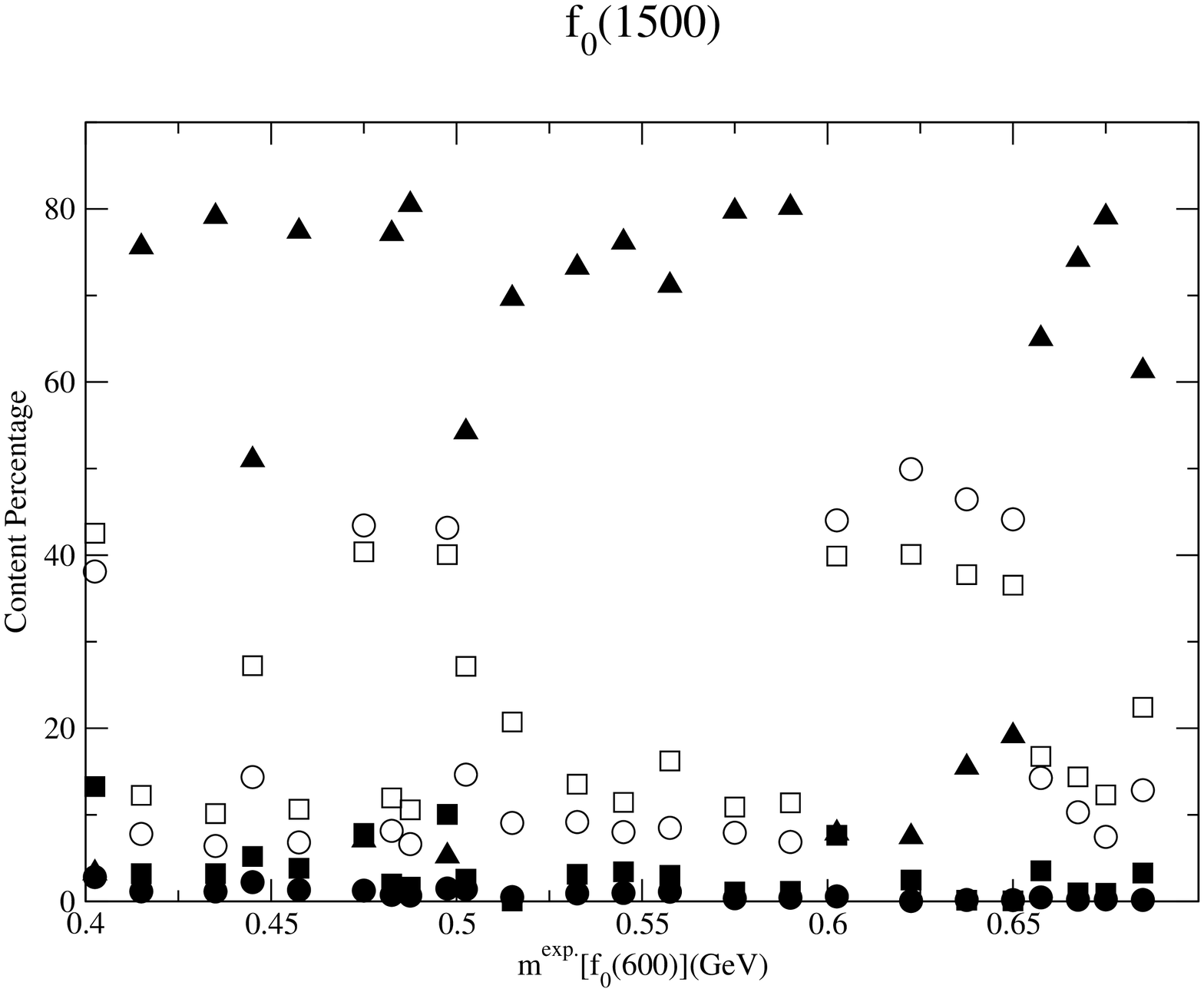}
\vskip .5cm
\epsfxsize=5cm
\epsfbox{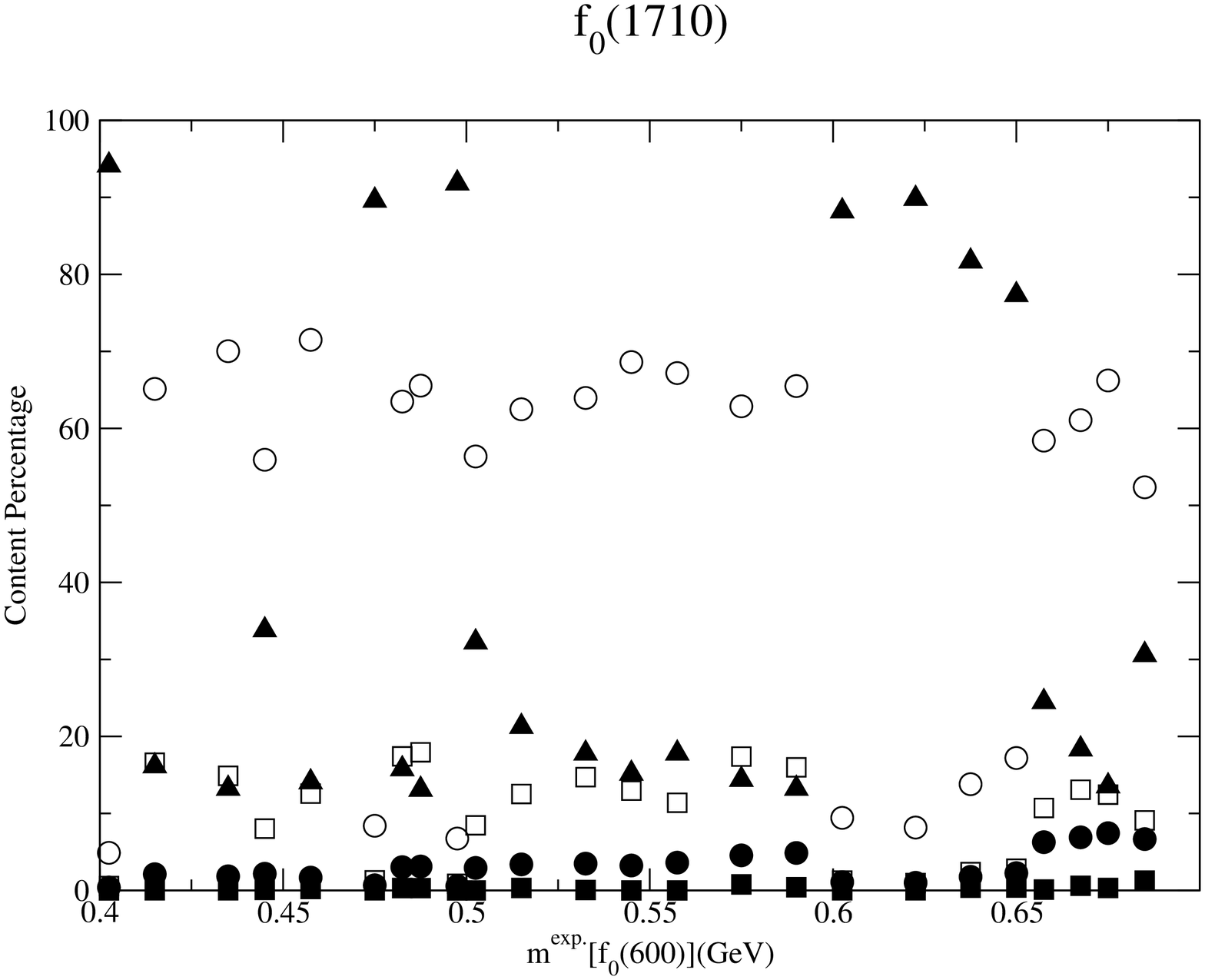}
\vspace*{8pt}
	\caption{Percentage of quark and glueball
components of the scalar states vs $m^{\rm exp.}
[f_0(600)]$:  ${\bar u}{\bar d} u d$ (filled circles),
$({\bar s}{\bar d} d s + {\bar s}{\bar u} u s) /\sqrt{2}$
(empty circles), ${\bar s}s$ (filled squares), $({\bar u}
u + {\bar d} d)/\sqrt{2}$ (empty squares), and glueball
(filled triangles).
	\label{F_comps_vs_ms}} 
\end{figure}

\begin{figure}[h]
\epsfxsize=5cm
\epsfbox{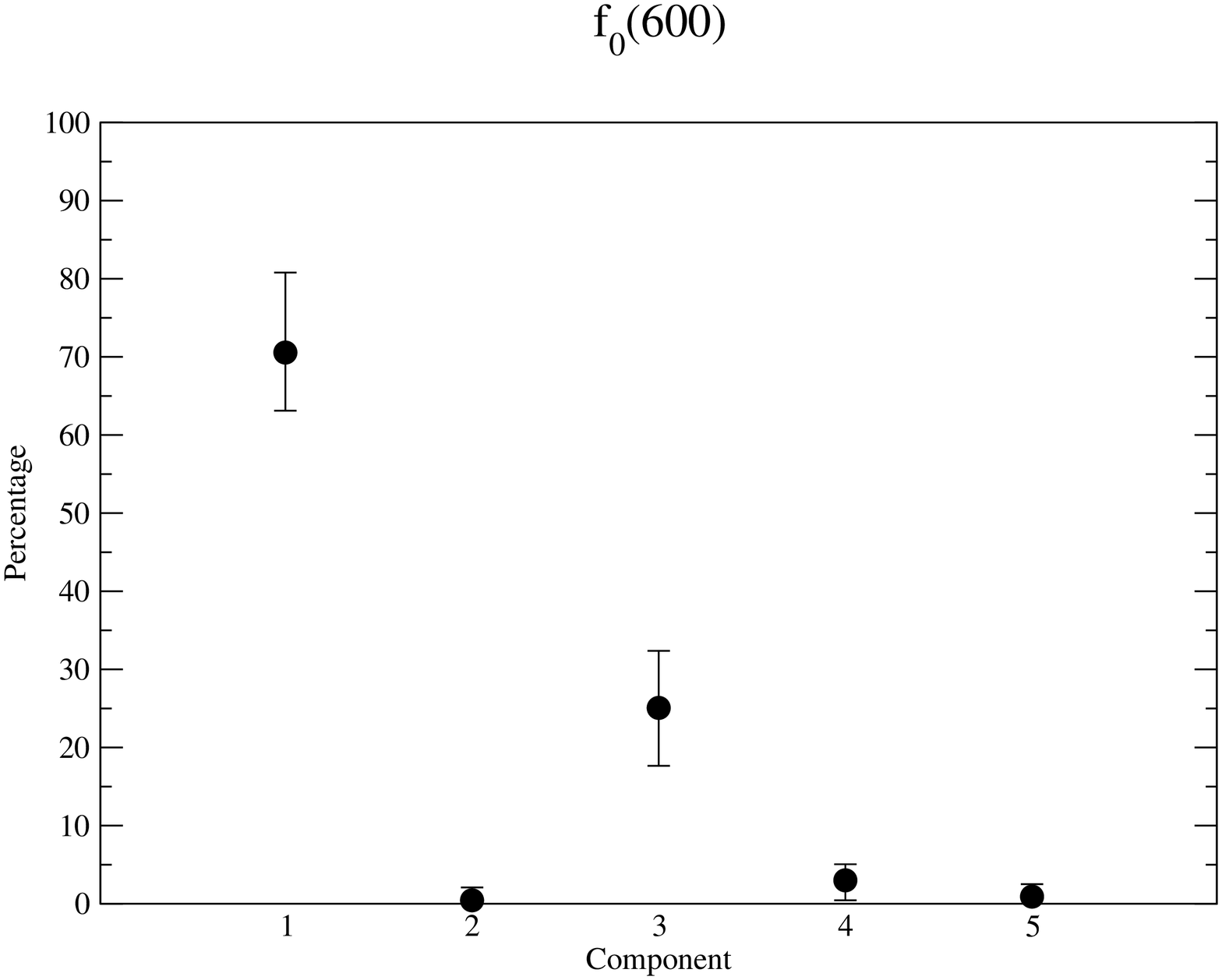}
\hskip .5cm
\epsfxsize=5cm
\epsfbox{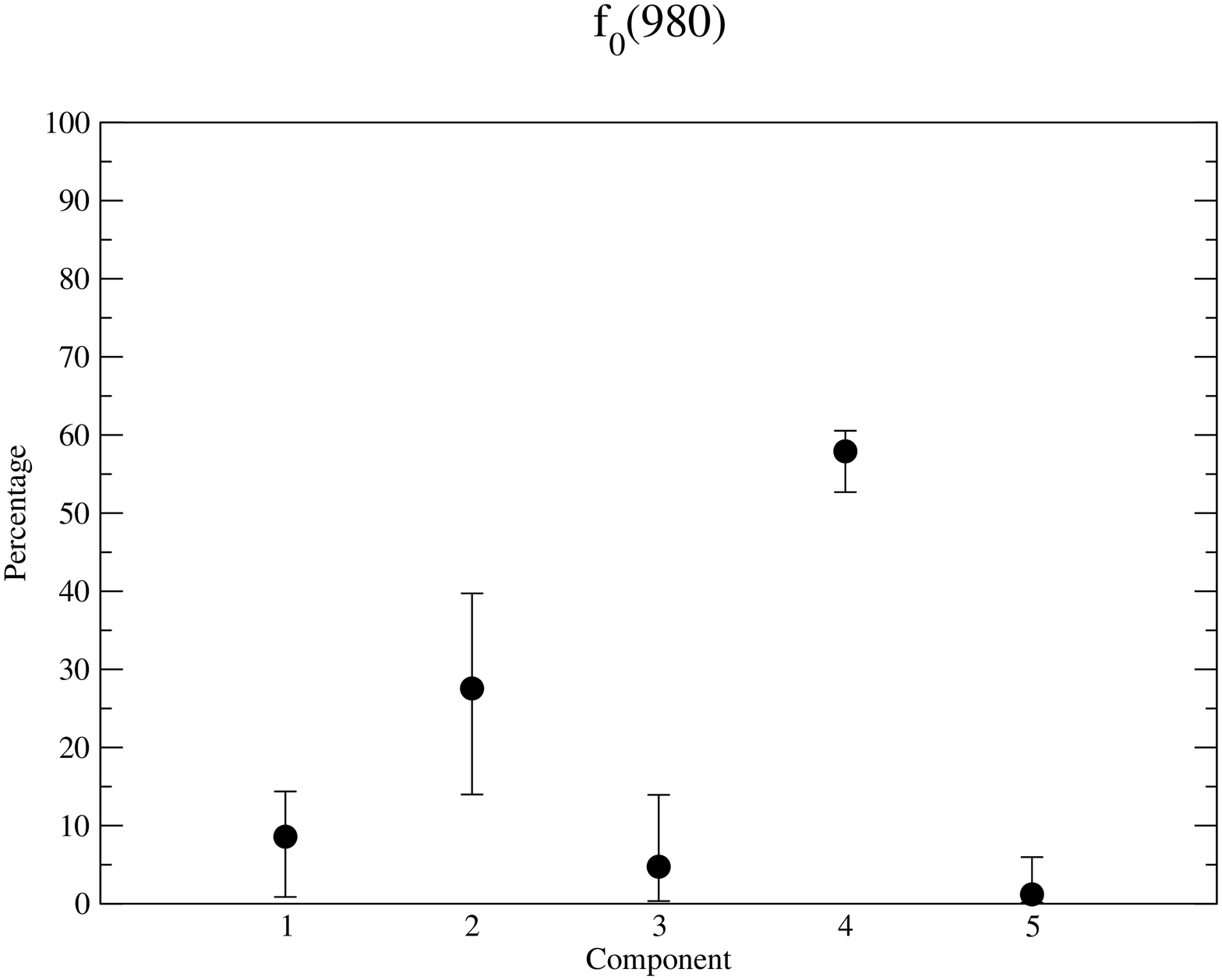}
\vskip .5cm
\epsfxsize=5cm
\epsfbox{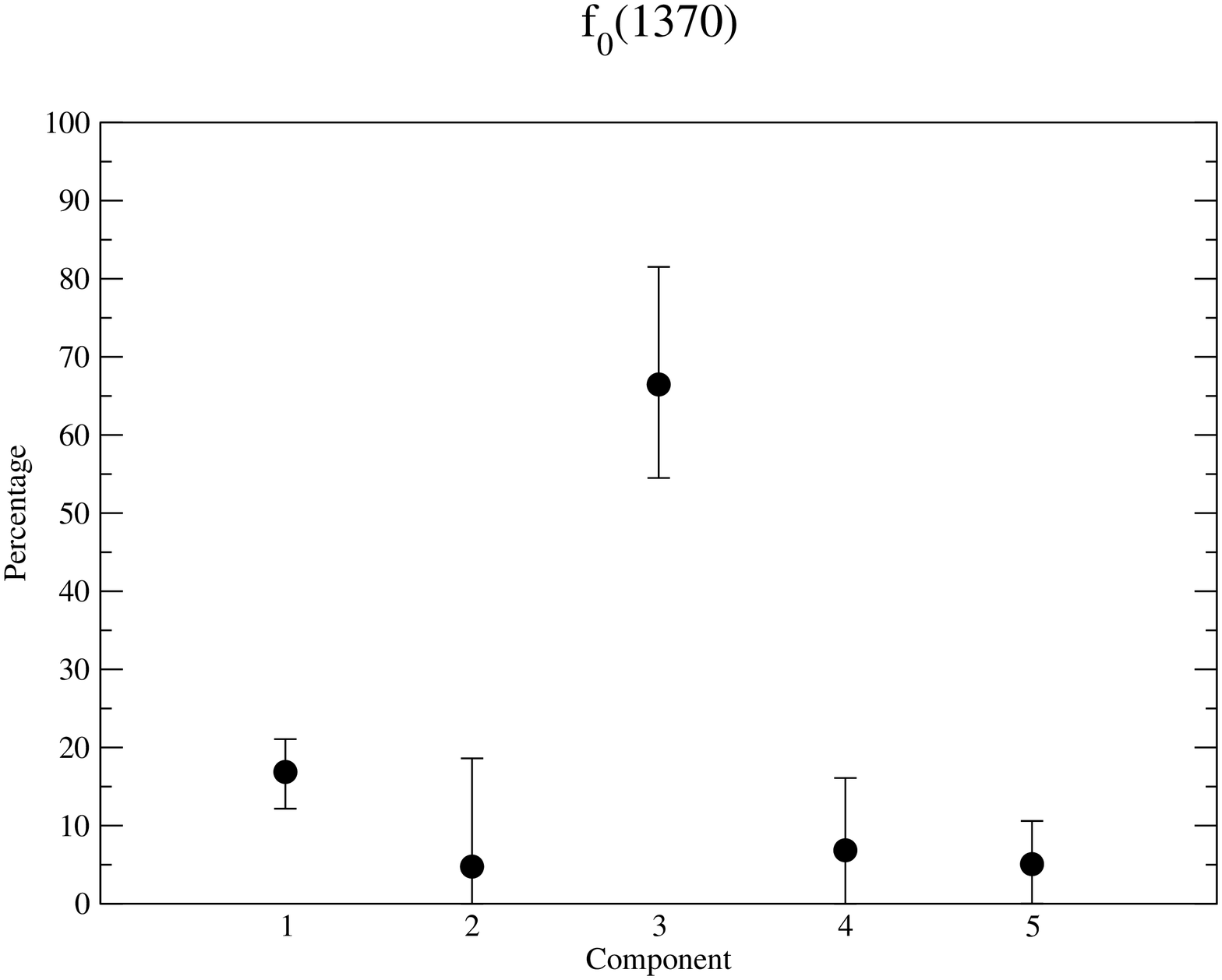}
\hskip .5cm
\epsfxsize=5cm
\epsfbox{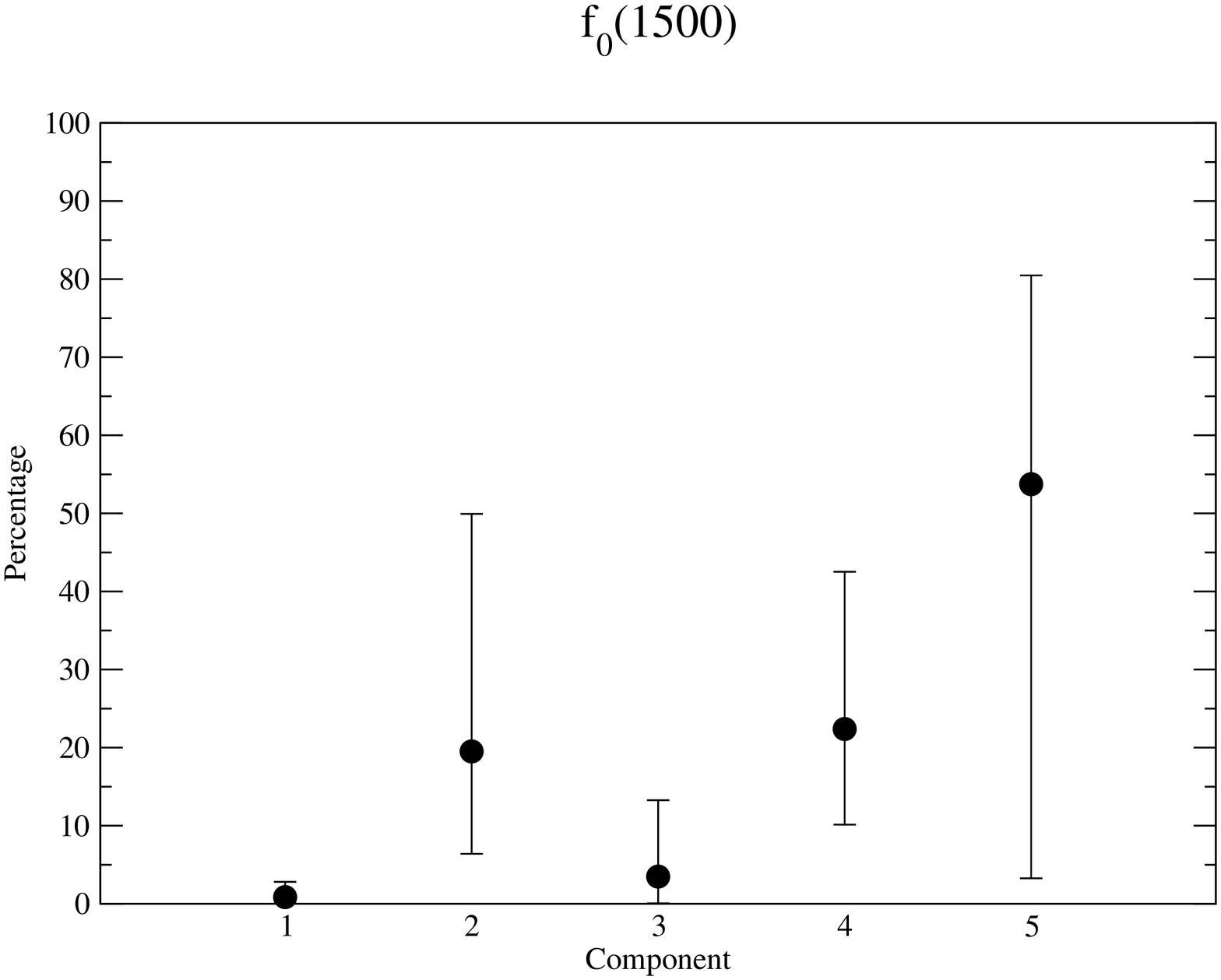}
\vskip .5cm
\epsfxsize=5cm
\epsfbox{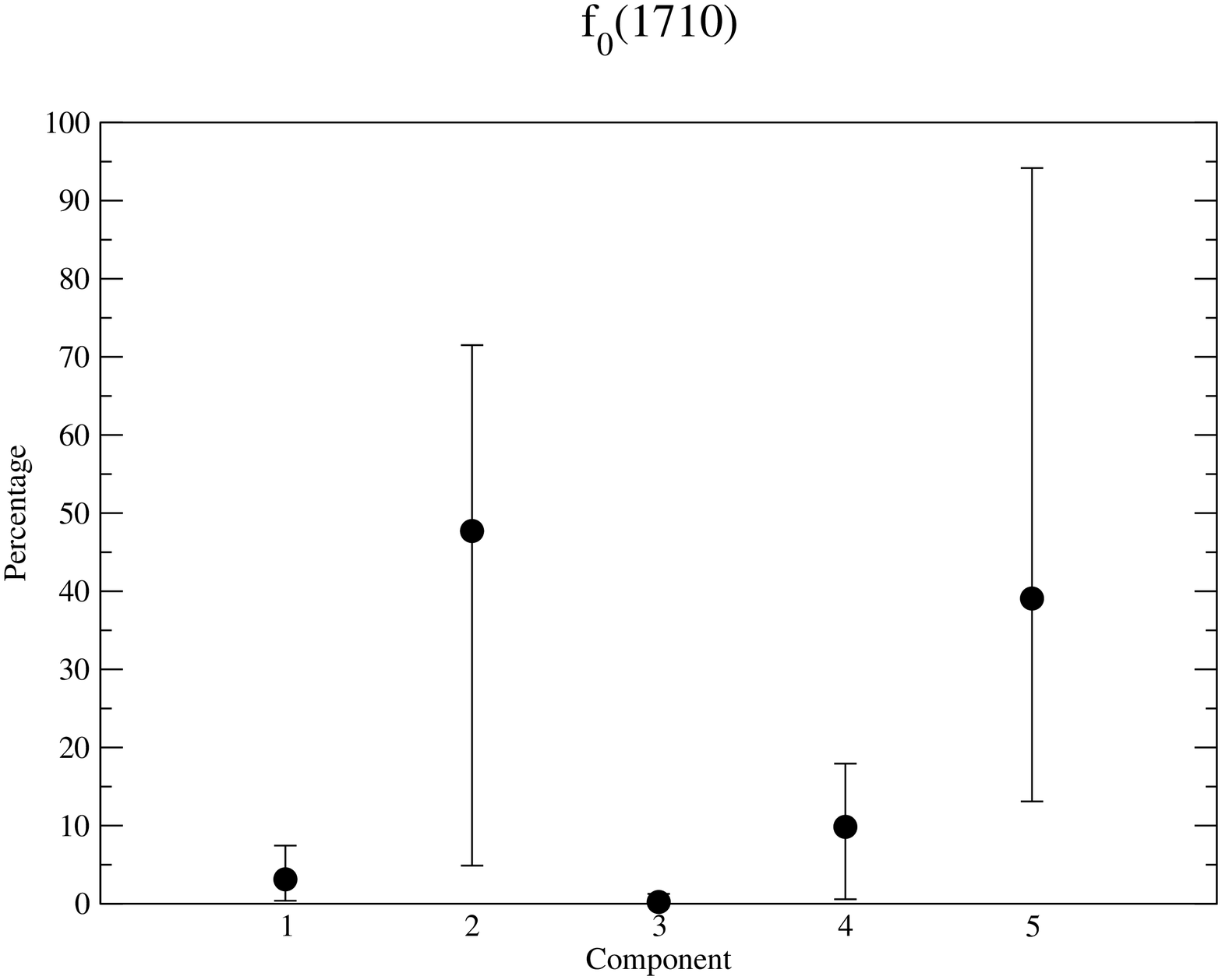}
\vspace*{8pt}
\caption{The effects of the mass uncertainties of the  
$f_0(600)$ and $f_0(1370)$ on the percentage of the quark and glueball 
components 
of the scalar states.   Components 1 to 5 respectively represent 
${\bar u}{\bar d} u d$, $({\bar s}{\bar d} d s + {\bar s}{\bar u} u s) 
/\sqrt{2}$, ${\bar s}s$,
$({\bar u} u + {\bar d} d)/\sqrt{2}$, and
glueball.     The dots represent the averaged values of each component and 
the error bars reflect the uncertainties of $m^{\exp.} [f_0(600)]$ and 
$m^{\rm exp.} [f_0(1370)]$.   The figures show that the components of 
the 
$f_0(600)$, 
$f_0(980)$ and $f_0(1370)$ are not very sensitive to these experimental 
uncertainties, but some of the components of the $f_0(1500)$ and 
$f_0(1710)$ (such as their glueball component) are significantly 
affected by such uncertainties.
\label{F_comps_error}} 
\end{figure}

At the local minima of Fig. \ref{F_ms_mf_plane}, we compute the rotation
matrix $K^{-1}$ (defined in Eq. (\ref{K_def})) which maps the quark and 
glueball bases to the physical
bases.  The results show that $K^{-1}$ is less sensitive to the mass of
the $f_0(600)$ and $f_0(1370)$.  For each scalar state, 
the quark and glueball 
components versus $m^{\rm exp.} [f_0(600)]$ are given in Fig.  
\ref{F_comps_vs_ms}, and the averaged components together with their 
variations are given in Fig. 
\ref{F_comps_error}.
Although the $m^{\rm
exp.} [f_0(600)]$ is still over a wide range of 400 to
700 MeV, we see in Fig. \ref{F_comps_vs_ms} that some of the components of
the physical state are qualitatively stable:  The admixtures of
$f_0(600)$, $f_0(980)$ and $f_0(1370)$, as well as some of the components
of the $f_0(1500)$ and $f_0(1710)$ are within a relatively small range of 
variation.   We see that the
$f_0(600)$ has a dominant ${\bar u}{\bar d} u d$
component and some ${\bar s}s$ content, and has almost
the reverse structure of the $f_0(1370)$;  the $f_0(980)$ is
dominantly a two quark non-strange state $({\bar u} u +
{\bar d} d)$ with a significant strange four-quark
content $({\bar s}{\bar d} d s + {\bar s}{\bar u} u s)$.   Although
sensitive to  the mass of the $f_0(600)$ and
$f_0(1370)$, we see that   
the $f_0(1500)$ contains a dominant glueball content with
some strange four-quark and non-strange two-qurak
admixtures;  and the $f_0(1710)$ is a dominant strange
four-quark state with a comparable glueball component.

\section{Summary and Conclusion}


We studied the $I=0$ scalar mesons below 2 GeV using a nonlinear chiral 
Lagrangian which is constrained by the mass
and the decay properties of the $I=1/2$ and $I=1$ scalar meson below 2 GeV
[$K_0^*(800)$, $K_0^*(1430)$, $a_0(980)$ and $a_0(1450)$]. This framework
provides an efficient approach for predicting the quark and glueball
content of the scalar mesons.  The main obstacle for a complete prediction
is the lack an accurate experimental input for the mass of $f_0(600)$ and
$f_0(1370)$.  Nevertheless, we showed that several relatively robust
conclusions can be made.   We showed that the $f_0(600)$, the $f_0(980)$ 
and the $f_0(1370)$ have a substantial admixture of two and four-quark 
components with a negligible glueball component.   The 
present model predicts that the $f_0(600)$ is dominantly a non-strange 
four-quark state; the $f_0(980)$ has a dominant non-strange two-quark 
component; and the $f_0(1370)$ has significant ${\bar s} s$ admixture.
We also showed that this model predicts that the $f_0(1500)$ and the 
$f_0(1710)$ have a considerable two and four quark admixtures, together 
with 
a dominant glueball component.    The current large uncertainties on the 
mass of $f_0(600)$ and $f_0(1370)$ do not allow an exact determination 
of the glueball components of the  $f_0(1500)$ and the
$f_0(1710)$, but it is 
qualitatively clear that the 
glueball components of these two states are quite large.    In addition 
the present model predicts that the glueball mass is in the range 
1.5$-$1.7 GeV
(Eq.({\ref{Gmass_range})).

The main 
theoretical improvement of the model involves inclusion of higher order 
mixing 
terms among nonets $N$ and $N'$ and the scalar glueball:
\begin{equation}
  {\rm Tr} \left( {\cal M} N N' \right)
               + {\rm Tr} \left( {\cal M} N' N \right)
\hskip .3cm , \hskip .3cm
{\rm Tr} ({\cal M} N) {\rm Tr} (N')
\hskip .3cm , \hskip .3cm
{\rm Tr} ({\cal M} N') {\rm Tr} (N)
\hskip .3cm , \hskip .3cm
e' G {\rm Tr} \left({\cal M}  N \right)
\hskip .3cm , \hskip .3cm
f' G {\rm Tr} \left({\cal M} N' \right)
\end{equation}
 
However, these terms both mix the quarks and glueballs as well as break
SU(3) symmetry, and therefore are of a more complex nature compared to the
terms used in this investigation (Eqs. (\ref{L_mix_I1})  and
(\ref{L_mix_I0})).  Investigation of such higher order mixing terms will
be left for future works.  It is also important to note that the results
presented here are not sensitive to the choice of $\gamma$ which 
determines
the mixing among the $I=1/2$ and $I=1$ mixings in Eq. (\ref{L_mix_I1}).
This is shown in Fig. \ref{F_comps_vs_g2} in which the components of all
$I=0$ states are plotted versus $\gamma^2$ and show that they are
relatively stable.  Also in Fig. \ref{F_baremasses} the bare masses are
plotted versus $\gamma^2$ in which we see that the expected ordering (i.e.
the lowest-lying four-quark nonet $N$ underlies a heavier two-quark nonet
$N'$ and a glueball) which is examined in ref.  \cite{Mec} using the
properties of the $I=1/2$ and $I=1$ scalar states, is not sensitive to the
choice of mixing parameter $\gamma$, providing further support for the
plausibility of the leading mixing terms considered in the present
investigation.

\begin{figure}[h]
\epsfxsize=5cm
\epsfbox{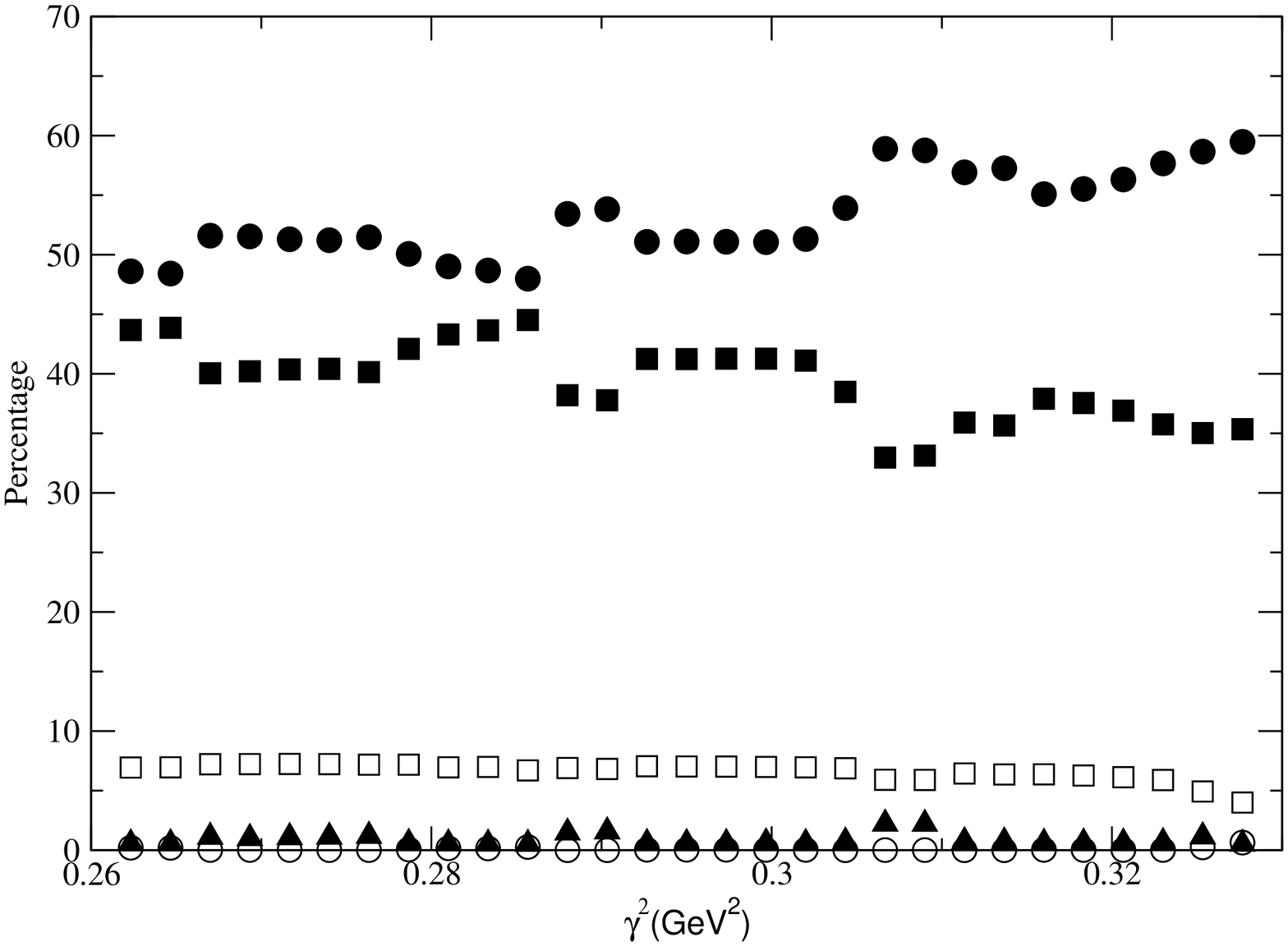}
\hskip .5cm
\epsfxsize=5cm
\epsfbox{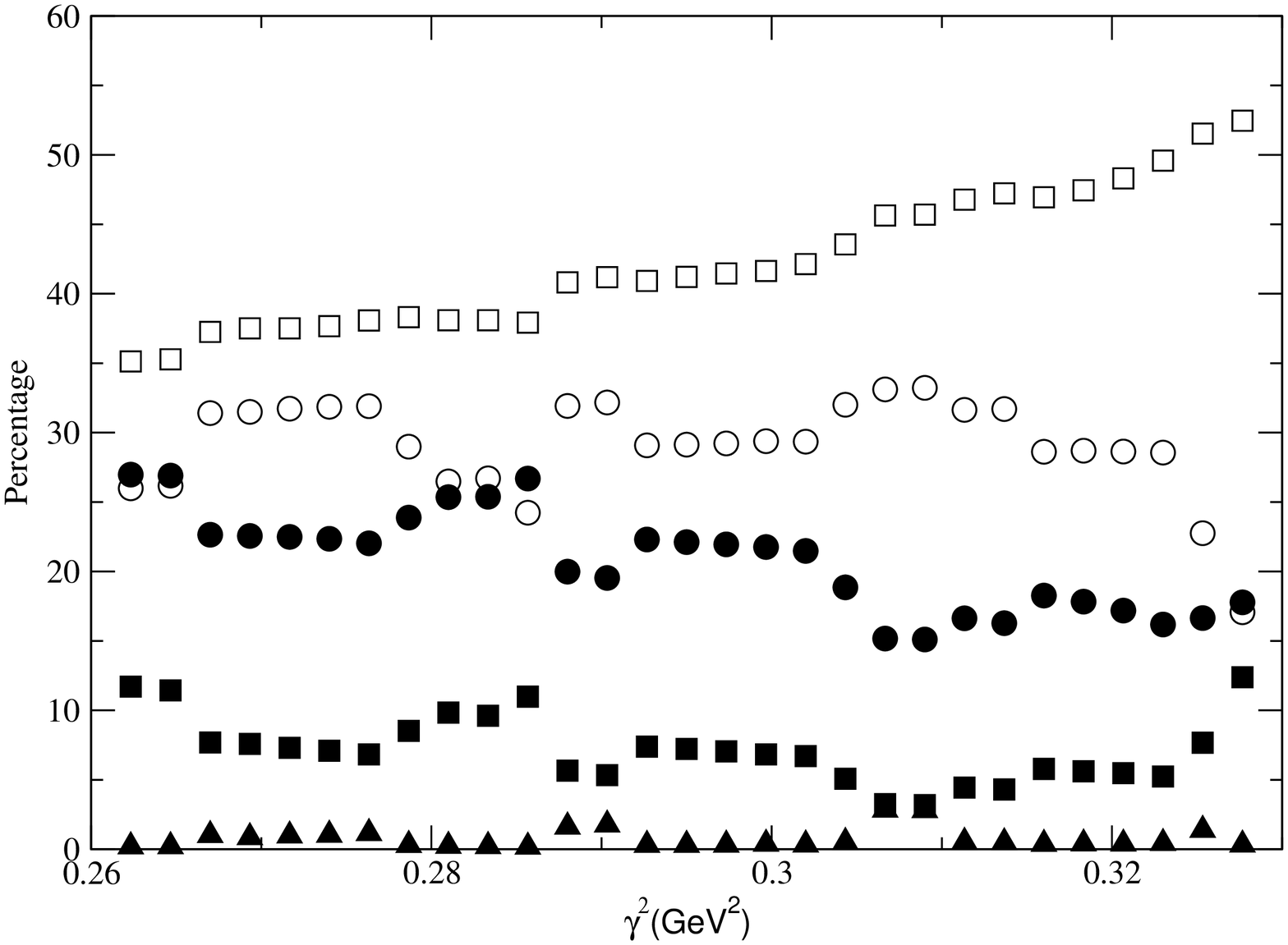}
\vskip .9cm
\epsfxsize=5cm
\epsfbox{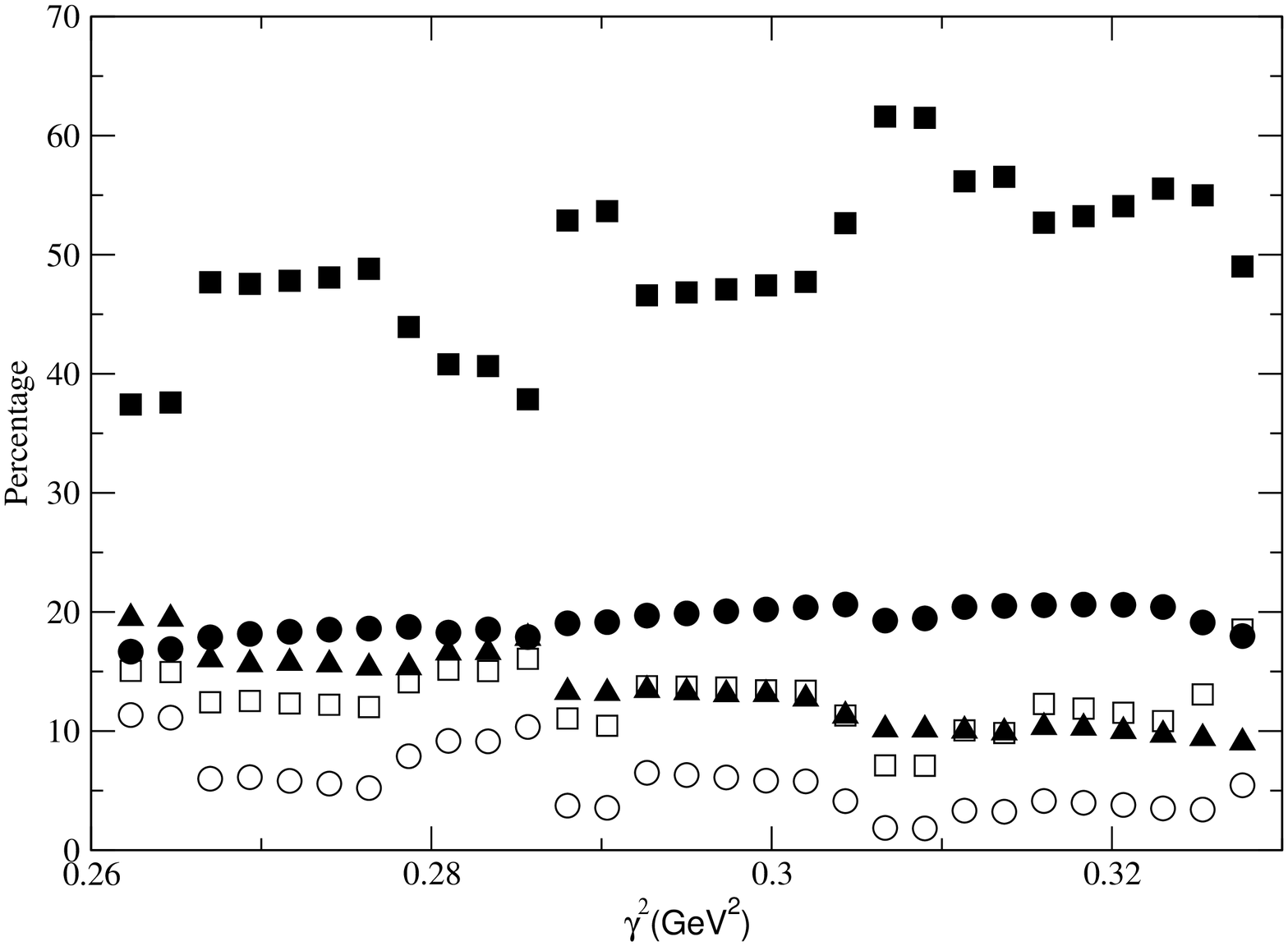}
\hskip .5cm
\epsfxsize=5cm
\epsfbox{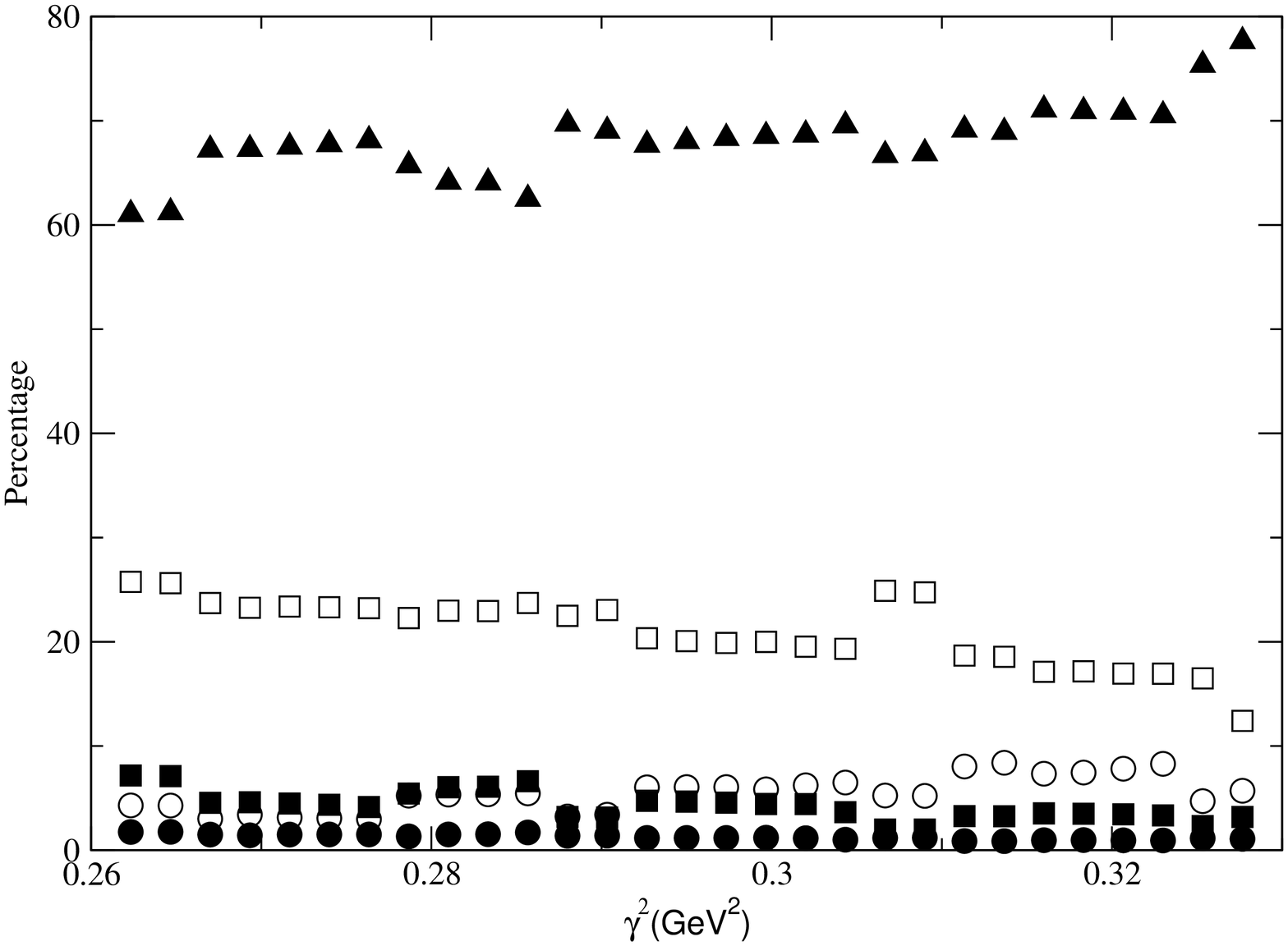}
\vskip .9cm
\epsfxsize=5cm
\epsfbox{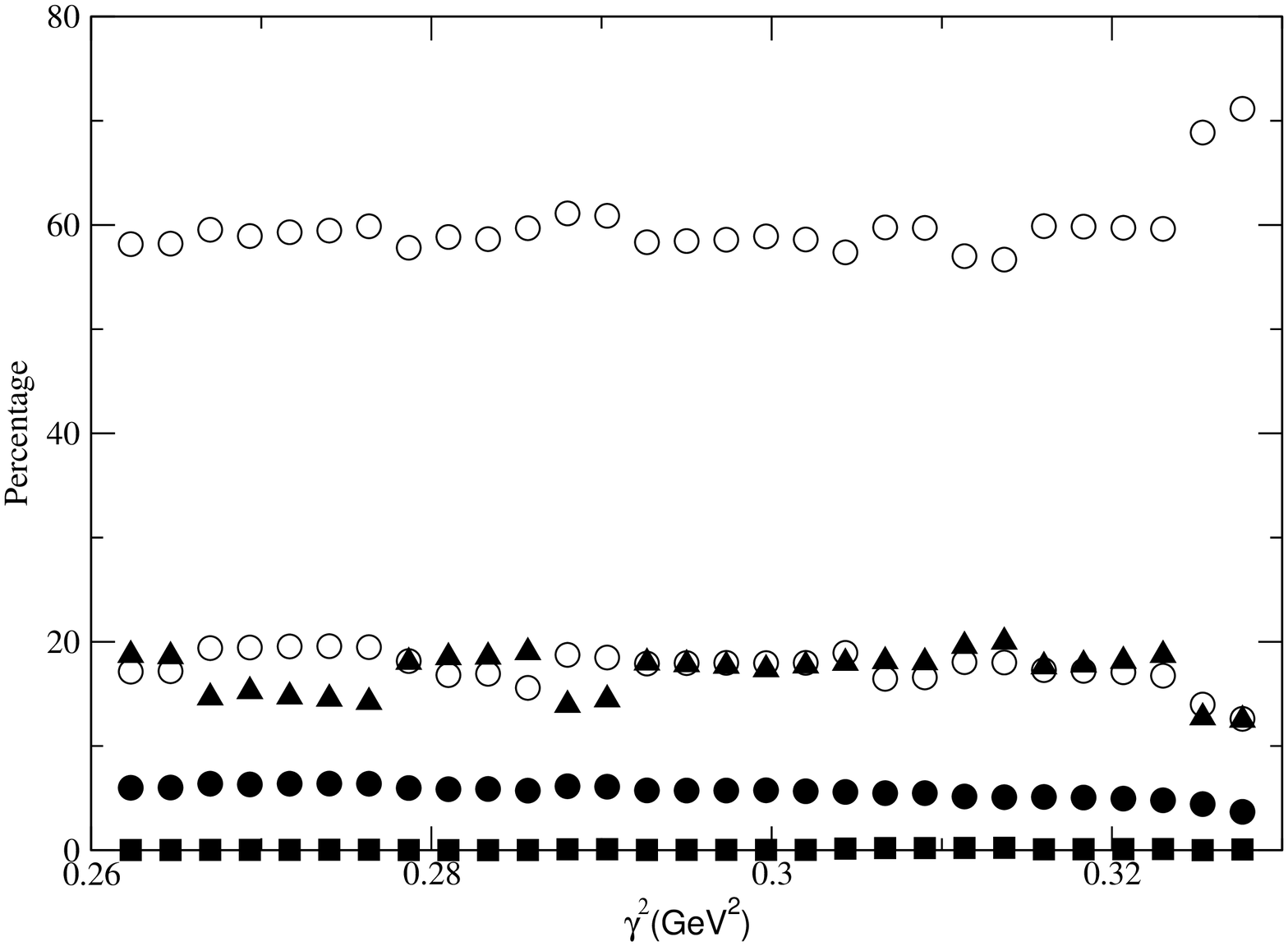}
\vspace*{8pt}
\caption{Percentage of quark and glueball components of the scalar 
states vs $\gamma^2$:  
${\bar u}{\bar d} u d$ (filled circles), 
$({\bar s}{\bar d} d s + {\bar s}{\bar u} u s) /\sqrt{2}$ (empty circles),
${\bar s}s$ (filled squares), 
$({\bar u} u + {\bar d} d)/\sqrt{2}$ (empty squares), and 
glueball (filled triangles).    The figures show that the components are 
not very sensitive to the mixing parameter $\gamma$.
\label{F_comps_vs_g2}} 
\end{figure}

\begin{figure}[h]
\epsfxsize=6cm
\epsfbox{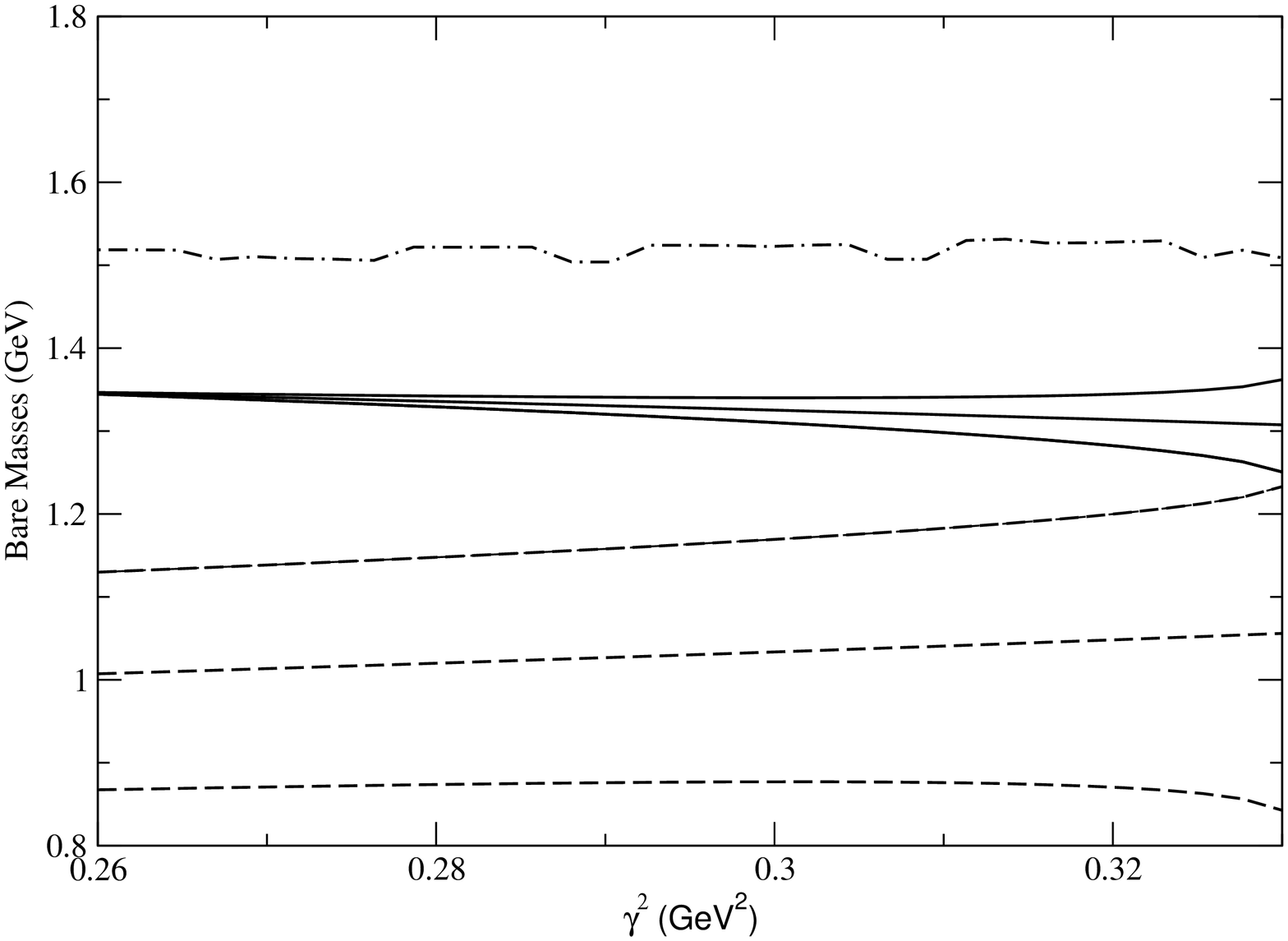} 
\vspace*{8pt} 
\caption{Dependence of the bare masses on $\gamma^2$.    The four-quark 
scalar nonet (dashed lines) lies below the quark-antiquark scalar nonet 
(solid lines) and a scalar glueball (dotted-dashed line). 
\label{F_baremasses}} \end{figure}

\section*{Acknowledgments}
The author wishes to thank M.R. Ahmady and J. Schechter for many helpful 
discussions.
This work has been supported in part by a 2005-2006 Crouse
Grant from the School of Arts and Sciences, SUNY
Institute of Technology.

\end{document}